\documentclass[prl,twocolumn,superscriptaddress,showpacs,aps]{revtex4-2}

\usepackage{graphicx,amsmath,amssymb}
\usepackage{float}

\newcommand{\ket}[1]{\lvert #1 \rangle}
\newcommand{\bra}[1]{\langle #1 \lvert}
\newcommand{\gc}{\Gamma_c}

\newcommand{\szj}{\hat{\sigma}^z_j}
\newcommand{\spj}{\hat{\sigma}^+_j}
\newcommand{\smj}{\hat{\sigma}^-_j}
\newcommand{\sxj}{\hat{\sigma}^x_j}
\newcommand{\syj}{\hat{\sigma}^y_j}

\newcommand{\mj}{\mathcal{J}}

\begin{document}
\title{Rugged mHz-Linewidth Superradiant Laser Driven by a Hot Atomic Beam}
\author{Haonan Liu}
\affiliation{JILA, National Institute of Standards and Technology, and University of Colorado, Boulder, Colorado 80309-0440, USA}
\author{Simon~B.~J{\"a}ger}
\affiliation{JILA, National Institute of Standards and Technology, and University of Colorado, Boulder, Colorado 80309-0440, USA}
\author{Xianquan Yu}
\affiliation{Centre for Quantum Technologies, Department of Physics, National University of Singapore, Singapore 117543}
\author{Steven Touzard}
\affiliation{Centre for Quantum Technologies, Department of Physics, National University of Singapore, Singapore 117543}
\author{Athreya Shankar}
\affiliation{JILA, National Institute of Standards and Technology, and University of Colorado, Boulder, Colorado 80309-0440, USA}
\author{Murray J. Holland}
\affiliation{JILA, National Institute of Standards and Technology, and University of Colorado, Boulder, Colorado 80309-0440, USA}
\author{Travis L. Nicholson}
\affiliation{Centre for Quantum Technologies, Department of Physics, National University of Singapore, Singapore 117543}

\date{\today}

\begin{abstract}
    We propose a new type of superradiant laser based on a hot atomic beam traversing an optical cavity. We show that the theoretical minimum linewidth and maximum power are competitive with the best ultracoherent clock lasers. Also, our system operates naturally in continuous wave mode, which has been elusive for superradiant lasers so far. Unlike existing ultracoherent lasers, our design is simple and rugged. This makes it a candidate for the first widely accessible ultracoherent laser, as well as the first to realize sought-after applications of ultracoherent lasers in challenging environments.
\end{abstract}

\maketitle

Ultracoherent light sources are the foundation of highly accurate atomic clocks~\cite{mcgrew2018,brewer2019}, measurements of the time variation of fundamental constants~\cite{huntemann2014,godun2014}, novel tests of relativity~\cite{sanner2019,kolkowitz2016}, and dark matter searches \cite{wcislo2018}. Traditionally, these sources have been generated with cavity stabilization, which involves locking lasers to highly stable optical cavities~\cite{hall2010}. Despite their incredible performance~\cite{robinson2019}, these systems are complex, challenging to improve upon, and perform poorly outside of controlled lab environments. However if cavity-stabilized lasers could be made rugged, they could be used for improved global positioning, deep space navigation~\cite{ely2018}, and new geophysical technology~\cite{bondarescu2015}.

Superradiant lasers~\cite{chen2009active, meiser2009prospects, Bohnet2012, Maier2014, roth2016, norcia2016coldSrLaser, Norcia2016mHz, laske2019, hotter2019, schaffer2020} are promising candidates for next-generation ultracoherent lasers~\cite{olson2019}. However, a continuous wave superradiant laser has not yet been demonstrated because of atomic heating in existing designs, which rely on ultracold atoms~\cite{norcia2016coldSrLaser}. Also, the use of ultracold atoms makes these systems complicated and ill suited to applications in the field.

Here we propose a new kind of superradiant laser built from a hot atomic beam traversing an optical cavity. We show that its theoretical minimum linewidth and maximum output power are competitive with the best ultracoherent lasers. Because of atomic phase synchronization, the phase of the output light is robust against decoherence arising from atomic motion, such as Doppler and transit time broadening. Furthermore, our system is naturally continuous wave, and it is inherently insensitive to effects that limit the best cavity-stabilized lasers~\cite{hall2010, robinson2019}, such as environmental noise and drift. The simplicity and ruggedness of the design make this system promising for applications in challenging real-world environments~\cite{takamoto2020,kolkowitz2016} and for packaging into commercial systems.

\begin{figure}[H]
    \centering
    \includegraphics[width=0.8\linewidth]{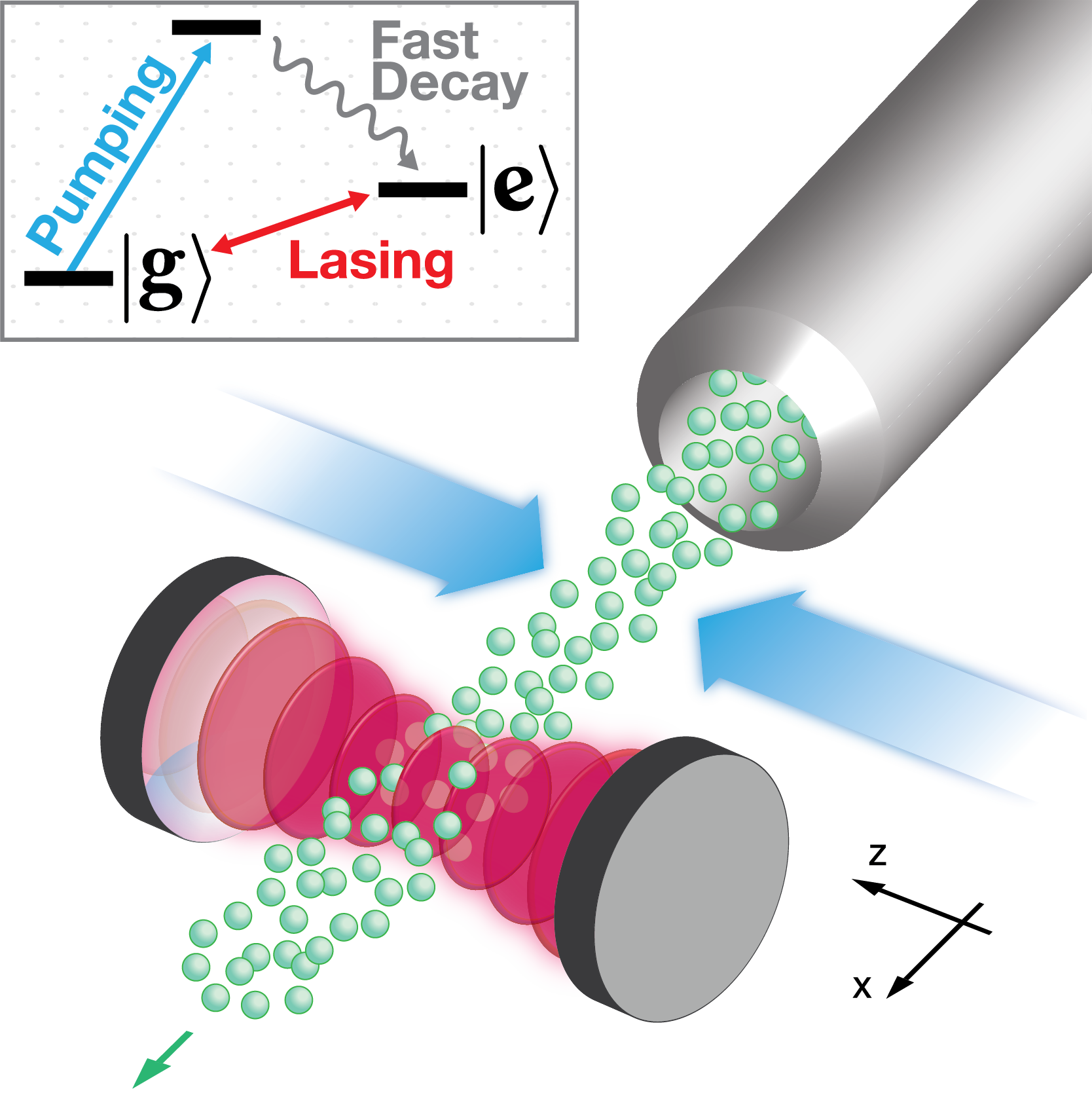}
    \caption{The superradiant beam laser. The atomic beam is generated from an effusive source, like a commercial effusion cell. After emerging from the source (upper right), the atoms are prepared by pumping lasers (blue arrows) in a metastable state prior to entering the cavity (lower left). Inset: The minimal atomic structure needed for the superradiant beam laser to operate. In this three-level scheme, atoms are rapidly prepared in a metastable state $\ket{e}$ by pumping (blue) on a broad transition. Lasing (red) occurs on the long-lived $\ket{g} \leftrightarrow \ket{e}$ transition. Real atomic systems may require more complex pumping schemes.}
    \label{fig: system}
\end{figure}

Our system consists of a dense atomic beam traveling through an optical cavity. We consider the case of all atoms having a uniform velocity in the $x$ direction (Fig.~\ref{fig: system}). In this work, we discuss the examples of $^{40}\mathrm{Ca}$ and $^{88}\mathrm{Sr}$, but our results apply equally well to many other alkaline-earth-like species. The mean intracavity atom number is $N\equiv\Phi\tau$ in steady state, where $\Phi$ is the number of atoms transiting the cavity mode per unit time, and $\tau$ is the transit time. The atoms in the beam are described by dipoles that are pumped into a metastable state (Fig.~\ref{fig: system}) before entering the cavity. The dipole transition frequency $\omega_a$ is taken to be near resonant with the frequency $\omega_c$ of a single cavity mode, where coupling of the dipoles and cavity is described by the Tavis-Cummings Hamiltonian $\hat{H}(t) =(\hbar g/2)\sum_{j} \eta[\mathbf{x}_j(t)](\hat{\sigma}_j^{+} \hat{a} + \hat{a}^{\dagger} \hat{\sigma}_j^{-})$. Here, the summation runs over all atoms in the beam, $\eta[\mathbf{x}_j(t)]$ is a cavity mode function evaluated at position $\mathbf{x}_j(t)$ of atom~$j$ at time $t$, and $g$ is the vacuum Rabi frequency at a cavity antinode. Furthermore, the atomic dipole raising and lowering operators are $\hat{\sigma}_j^+ = \left(\hat{\sigma}_j^-\right)^\dagger = \ket{e}_j\bra{g}_j$, where $\ket{g}$ and $\ket{e}$ are the atomic ground and excited states, respectively, and the photon annihilation and creation operators of the cavity field mode are $\hat{a}$ and $\hat{a}^{\dagger}$. Besides this Hamiltonian that couples the atoms and cavity, our model includes photon loss through a cavity mirror with rate $\kappa$.

We consider the bad cavity regime, which occurs when $\kappa$ is much larger than the transit time broadening $1/\tau$, the collective coupling $\sqrt{N}g$, and the Doppler width $\delta_D=k\Delta v_z$. Here $k=2\pi/\lambda$, $\lambda$ is the optical wavelength, and $\Delta v_z$ is the single-atom velocity width along the cavity axis. In this regime, the light field is rigidly anchored to the collective atomic dipole, so that the cavity degrees of freedom can be adiabatically eliminated as $\hat{a}\approx -ig \hat{J}^-/\kappa$. The operator $\hat{J}^-=\sum_j\eta(\mathbf{x}_j)\smj$ is the collective dipole, which is the sum of the individual atomic dipoles interacting with the cavity mode. New atomic dipoles entering the cavity synchronize with the existing collective dipole due to the atom-cavity interaction~\cite{xu2014synch}. Since there is a large number of atoms in the cavity mode, the true operator equations are well approximated by stochastic differential equations for their complex amplitude equivalents~\cite{SM};
\begin{align}
    \frac{ds_j^x}{dt} &= \frac{\gc}{2}\eta_j\left[\mj^x s_j^z - \eta_js_j^x (s_j^z + 1)\right] - \sqrt{\gc} \eta_js_j^z \xi^p,
    \label{sx}\\
    \frac{ds_j^y}{dt} &= \frac{\gc}{2}\eta_j\left[\mj^y s_j^z - \eta_js_j^y (s_j^z + 1)\right] + \sqrt{\gc} \eta_js_j^z \xi^q,
    \label{sy}\\
    \frac{ds_j^z}{dt} &= - \frac{\gc}{2}\eta_j\left\{\mj^x s_j^x + \mj^y s_j^y - \eta_j\left[\left({s_j^x}\right)^2 + \left({s_j^y}\right)^2\right]\right\} \nonumber\\
                       &- \gc \eta_j^2(s_j^z + 1)+ \sqrt{\gc}\eta_j \left(s_j^x \xi^p - s_j^y \xi^q\right).
    \label{sz}
\end{align}
Here $s_j^x$, $s_j^y$, and $s_j^z$ are the $c$-number pseudospin variables that correspond to $\sxj=\smj+\spj$, $\syj=i(\smj-\spj)$, and $\szj=\spj\smj-\smj\spj$. Similarly, $\mj^x$ and $\mj^y$ represent the operators $\hat{J}^x=\hat{J}^-+\hat{J}^+$ and $\hat{J}^y=i(\hat{J}^--\hat{J}^+)$. We have defined $\Gamma_c=\mathcal{C}\gamma$, where $\mathcal{C}=g^2/(\kappa\gamma)$ is the cavity cooperativity and $\gamma$ is the free-space spontaneous emission rate. We use the shorthand $\eta_j=\eta[\mathbf{x}_j(t)]$ and model the cavity mode by $\eta(\mathbf{x})=[\Theta(x+w)-\Theta(x-w)]\cos(kz)$, where $\Theta(x)$ is the Heaviside step function and $w$ is the cavity beam waist. Spontaneous emission into free space is neglected in Eqs.~\eqref{sx}--\eqref{sz} because the collective lifetime is much shorter than the spontaneous lifetime~\cite{footnote1, meiser2009prospects, meiser2010ssLaser, xu2014synch}. Along the cavity axis, the atoms are randomly assigned a velocity drawn from a Maxwell-Boltzmann distribution at a given temperature. Cavity shot noise is denoted by the stochastic noise variables $\xi^q$ and $\xi^p$, which have zero mean and are delta correlated as $\langle \xi^{a}(t)\xi^{b}(t')\rangle=\delta_{ab}\delta(t-t')$, $a,b\in\{q,p\}$. Each atom enters the cavity with $s^z_j=1$, and projection noise is included by choosing random (and independent) values $+1$ or $-1$ for $s^x_j$ and $s^y_j$~\cite{schachenmayer2015monteCarlo, SM}.

\begin{figure*}
    \centering
    \includegraphics[width=0.9\textwidth]{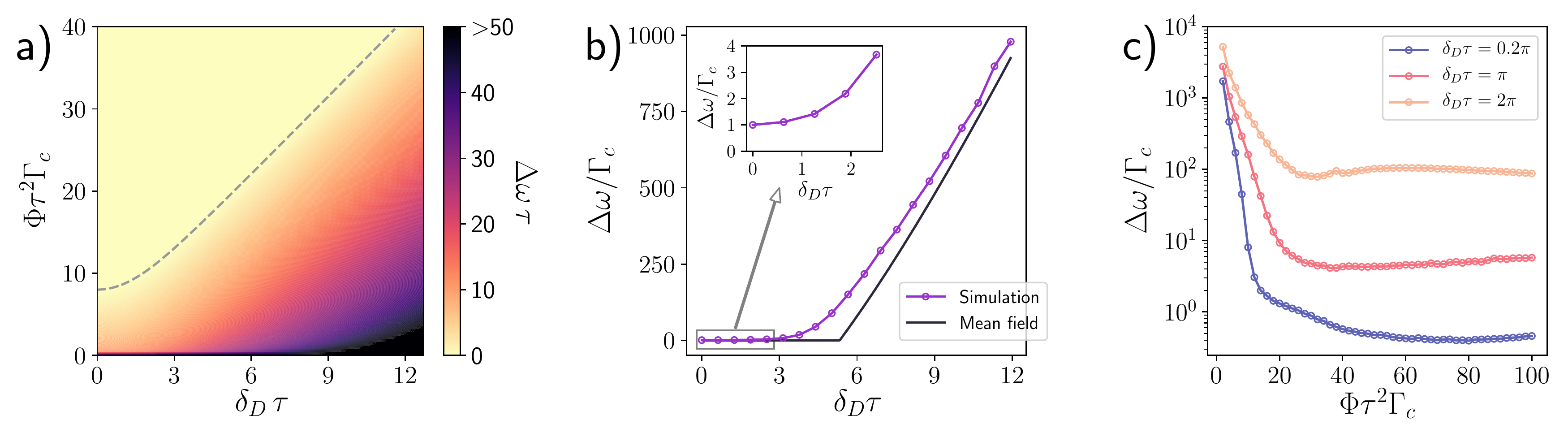}
    \caption{(a) Mean-field calculations of the linewidth in units of the transit time broadening $1/\tau$, as a function of the Doppler width $\delta_D\tau$ and $\Phi\tau^2\Gamma_c$. Here $\Phi\tau^2\Gamma_c$ is the number of collective lifetimes that elapse during the transit time $\tau$. The black dashed line is the phase transition threshold for steady-state superradiance, above which mean-field calculations predict a zero linewidth. (b) The linewidth in units of $\Gamma_c$ as a function of the Doppler width for $\Phi\tau^2\Gamma_c=20$. The markers are simulation results using Eqs.~\eqref{sx}--\eqref{sz} with $\Phi=1000/\tau$ and $\Gamma_c=0.02/\tau$. For every data point, we calculated 100 trajectories each with a simulation time of $T=2000\tau$. This numerical simulation is compared with the mean-field theory, which is analytic. Inset: Below the phase transition, simulations show an ultranarrow linewidth of order $\Gamma_c$, which is 50 times smaller than transit time broadening for these simulation parameters. (c) Simulation results of the linewidth in units of $\Gamma_c$ as a function of $\Phi\tau^2\Gamma_c$. For every data point, we calculated 100 trajectories each with a simulation time of $T=200\tau$ and $\Phi=500/\tau$.}
    \label{fig: linewidth}
\end{figure*}

Typically, resonance widths in hot gases of atoms are dominated by Doppler and transit time broadening. Although our system is based on a hot gas, these broadening mechanisms vanish when the collective linewidth $N\Gamma_c$ is much greater than $\delta_D$ and $1/\tau$. The collective linewidth $N\Gamma_c$ is the rate for an atom to spontaneously emit into the cavity in the presence of other atoms. The principal features of this model can be obtained by dropping the noise terms in Eqs.~\eqref{sx}--\eqref{sz}, corresponding to a mean-field solution that is simple enough to be solved analytically and allows us to classify different phases of emission. The form of the solution for the laser linewidth $\Delta \omega$ is determined by two independent parameters, the first being $\delta_{D}$ and the second being $\Phi\tau^2\Gamma_c=\tau/(N\Gamma_c)^{-1}$, which is the number of collective lifetimes that elapse during $\tau$. In general, we observe a phase transition from broad linewidth emission to superradiant emission with an ultranarrow linewidth [Fig.~\ref{fig: linewidth}(a)]~\cite{SM}. Specifically, for large $\delta_{D}\tau$, the transition threshold is governed by the Doppler width, whereas for small $\delta_{D}\tau$, transit time broadening determines the regime of superradiant emission. The latter is evident because there is no superradiant emission for $\Phi\tau^2\Gamma_c<8$ even in the absence of Doppler broadening ($\delta_{D}\tau \ll 1$). This is because unsynchronized atoms are introduced to the cavity so rapidly that the collective dipole does not establish.

The mean-field analysis predicts an unphysical zero linewidth in the superradiant regime because it neglects quantum noise. In reality, vacuum fluctuations entering the cavity and quantum fluctuations in the atomic dipole components cause phase diffusion, resulting in a nonvanishing linewidth. To determine this linewidth we simulate Eqs.~\eqref{sx}--\eqref{sz} with noise terms included for $\Phi\tau^2\Gamma_c = 20$. The mean-field theory and $c$-number simulations agree outside the superradiant regime, whereas inside the superradiant regime only the $c$-number simulations predict a nonvanishing linewidth. Here the minimum achievable linewidth is $\Gamma_c$ [Fig.~\ref{fig: linewidth}(b) inset], which is much smaller than $1/\tau$, implying that our system is robust against single-atom transit time broadening. In other words, the collective atomic dipole stores the optical phase for much longer than the time any individual atom spends in the cavity. 

To see how the minimum linewidth in the superradiant phase varies with $\delta_D$, we run simulations with three Doppler widths [Fig.~\ref{fig: linewidth}(c)]. For $\delta_D\tau=\pi$, the linewidth can be brought down to several $\Gamma_c$, and for $\delta_D\tau=0.2\pi$, the linewidth is $\Gamma_c$. These numbers elucidate that narrow-linewidth superradiant emission occurs when the atoms are flying through the cavity so quickly that they move less than $\lambda/2$ along the cavity axis during $\tau$.

To understand the scale of these quantities, we evaluate numerical values for the $^3P_1 \rightarrow {}^1S_0$, $\gamma=2\pi \times 400\, \mathrm{Hz}$ transition in $^{40}\mathrm{Ca}$. We take the velocity in the $x$ direction to be that of $\mathrm{Ca}$ atoms from an effusion cell operating at $\sim 800 \, ^{\circ}\mathrm{C}$. We also consider the case where $\Phi \sim 10^{14} \, \mathrm{atoms/s}$ and the atomic beam is laser cooled in the transverse direction to $\Delta v_z \simeq 0.41 \, \mathrm{m/s}$, corresponding to the $\delta_D \tau = \pi$ curve in Fig.~\ref{fig: linewidth}(c). Considering a simple cavity with straightforward dimensions (a finesse of 20, a cavity length of $3 \, \mathrm{cm}$, and beam waist $w = 300 \, \mathrm{\mu m}$), we calculated a minimum linewidth of order $10 \,\mathrm{mHz}$~\cite{SM}, competitive with the best stable lasers to date~\cite{Milner_2019}. A similar analysis based on the ${}^{88}$ $\mathrm{Sr}$ intercombination transition yields a minimum linewidth of order 100 mHz~\cite{SM}. Therefore, ultracoherent light can be extracted from a hot atomic beam with a significant Doppler width, which implies that ultracold atoms may not be required to achieve narrow linewidth superradiant laser emission.

\begin{figure}
    \centering
    \includegraphics[width=\linewidth]{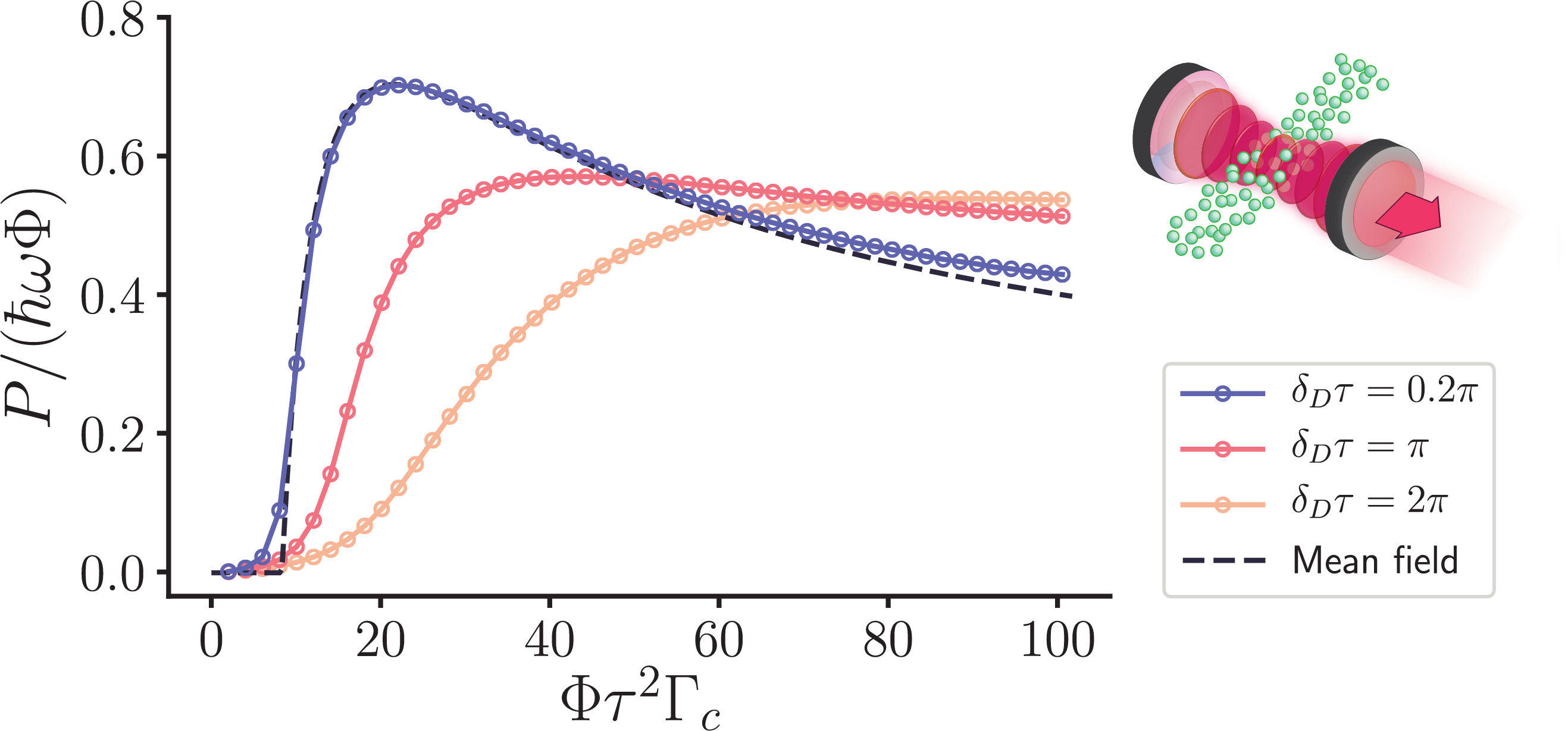}
    \caption{The output power of the superradiant beam laser. The markers are $c$-number simulation results. For every data point, we calculated 100 trajectories each with a simulation time of $T=200\tau$ and $\Phi=500/\tau$. For $\delta_D\tau=0.2\pi$, both the mean-field and simulation results peak at $\Phi\tau^2\Gamma_c \approx 20$ with $P = 0.7\hbar\omega\Phi$.}
    \label{fig: intensity}
\end{figure}

We now turn our attention to the laser output power~$P$. While individual atoms would rarely emit into the cavity mode during their passage, the emission rate is greatly enhanced by collective effects. This enhanced rate leads to a $N^2$ power scaling~\cite{SM, meiser2009prospects, meiser2010ssLaser}, which is a principal feature of superradiant emission. Determining $P$ from both the mean-field and $c$-number simulation approaches, we find good agreement between the two when $\delta_D\tau$ is comparable to (or below) $0.2\pi$ (Fig.~\ref{fig: intensity}). For Doppler widths in this regime and for $\Phi\tau^2\Gamma_c = 2\pi^2 \approx 20$, $P$ achieves its maximum value of $0.7\hbar\omega\Phi$, where $\omega$ is the center frequency of the output field. Physically this corresponds to each atom emitting an average of $0.7$ photons into the cavity mode. Furthermore, we find that the emitted light is second-order coherent by calculating $g^{(2)}(0)\approx1$ (as shown in the Supplemental Material~\cite{SM}). Together, Fig.~\ref{fig: linewidth}(c) and Fig.~\ref{fig: intensity} show that the maximum power and a linewidth of order $\Gamma_c$ can be simultaneously achieved when $\Phi\tau^2\Gamma_c \approx 20$ and $\delta_D\tau \lesssim 0.2\pi$.

For the $^{40}\mathrm{Ca}$ example mentioned above, we find that $P \approx 0.1\,  \mathrm{mW}$ at a linewidth of $40\, \mathrm{mHz}$. For ${}^{88}\mathrm{Sr}$, $P = 2.5 \, \mathrm{\mu W}$ at a linewidth of 150 mHz. Significantly, these powers should be sufficient for use with standard laser technology. In contrast the previously considered cold atom version of the superradiant laser has orders of magnitude weaker power, restricting its use to specialized equipment~\cite{meiser2009prospects}. The power $P$ is greater in the superradiant beam laser because it has the potential for a much larger intracavity atom number than cold atom systems, where particle numbers have been limited by intrinsic inefficiencies in ultracold gas preparation techniques.

\begin{figure}
    \centering
    \includegraphics[width=\linewidth]{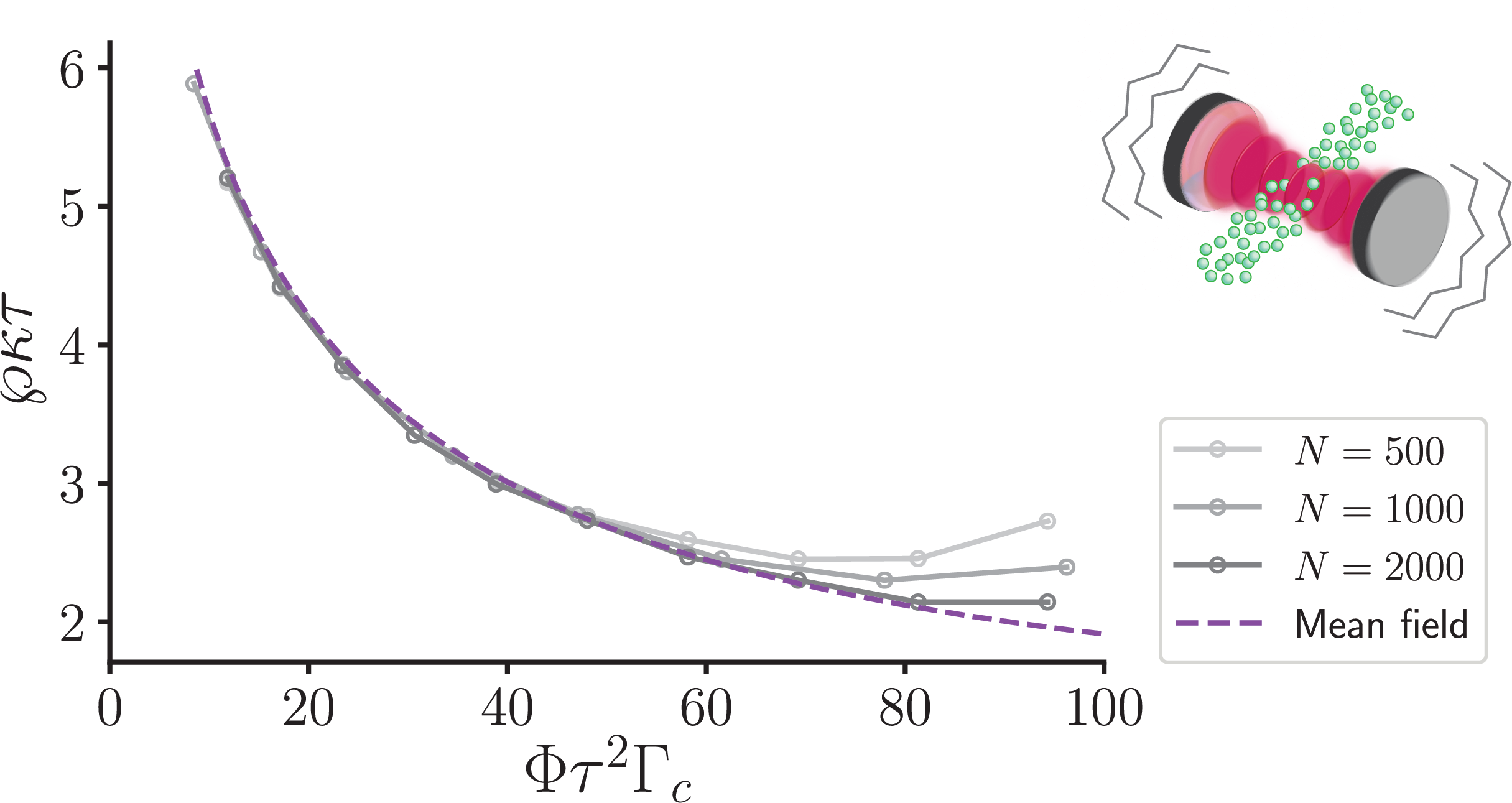}
    \caption{The cavity pulling coefficient $\wp$ at $\delta_D\tau=0.2\pi$. A small cavity pulling makes the laser frequency insensitive to environmental noise, such as vibrations. The markers are $c$-number simulation results with $\kappa=1000/\tau$ and $\omega_c-\omega_a=100/\tau$. For every data point, we calculated 100 trajectories each with a simulation time of $T=100\tau$. As $N$ increases, the simulation results approach the mean-field calculation.}
    \label{fig: pulling}
\end{figure}

In addition to its relatively large output power and insensitivity to Doppler and transit time broadening, this design is robust against environmental noise. This noise causes cavity length fluctuations, which manifest as cavity resonance frequency noise that dominates the linewidths of cavity-stabilized narrow-linewidth lasers~\cite{martin2012}. For these lasers, the frequency noise on the laser output field is equal to the environmental noise in the cavity resonance frequency. However, in a superradiant laser, phase information is stored primarily in the atomic medium, which makes the phase rigid against cavity resonance fluctuations; therefore, these fluctuations are written onto the laser output frequency with a strong suppression factor. This factor is the cavity pulling coefficient~\cite{Bohnet2012}, defined as $\wp = (\omega-\omega_a)/(\omega_c-\omega_a)$, which is the fractional change in the laser frequency when the cavity resonance fluctuates with respect to the atomic transition. Using mean-field theory, we analytically find that $\wp \propto 1/(\kappa\tau)$, which is the ratio of the cavity photon lifetime to the atom transit time. A value of $\kappa\tau = 1000$ can be achieved with standard optics~\cite{SM}, resulting in $\wp \approx 0.004$ for $\Phi\tau^2\Gamma_c=20$ (see Fig.~\ref{fig: pulling}).

This small $\wp$ makes our design robust against environmental noise sources that limit linewidths of cavity-stabilized lasers. The most common examples are vibration noise~\cite{filler1988}, thermal Brownian noise~\cite{martin2012}, and slow drift in the cavity length~\cite{SM}. The response of cavity resonance frequency to vibration noise is characterized by the acceleration sensitivity $K$. For the superradiant beam laser, the laser output frequency has an effective acceleration sensitivity $\wp K$. If our design uses a simple V-block cavity with no regard for the vibration isolation found in cutting-edge stable lasers, it would have an acceleration sensitivity of $\wp K \sim 10^{-13}/(\mathrm{m/s^2})$ \cite{SM}. Meanwhile, the acceleration sensitivity of the best cavity-stabilized laser to date is of the same order, i.e., $K \sim 10^{-13}/(\mathrm{m/s^2})$~\cite{robinson2019}.

Thermal Brownian noise causes cavity resonance fluctuations that scale as $1/L$, where $L$ is the cavity length. To suppress this effect, stabilization cavities have been made as long as half a meter~\cite{hafner2015}. For the superradiant beam laser, the amplitude of thermal noise behaves according to the effective cavity length $L/\wp$. This means that the output frequency of a beam laser based on a compact $L = 3 \, \mathrm{cm}$ cavity has the thermal noise of a $7.5 \, \mathrm{m}$ cavity. Furthermore, slow thermal drift is a practical challenge for cavity-stabilized lasers. The superradiant beam laser has an effective coefficient of thermal expansion (CTE) of $\wp\alpha$, where $\alpha$ is the CTE of the bare cavity. This means that a beam laser based on Invar (an inexpensive and easy-to-machine material) with modest temperature control would have a drift rate similar to that of an ultrastable cavity based on highly temperature-stabilized ultralow expansion glass~\cite{SM}.

Our model is intended to be a simple and clear treatment that correctly reproduces the laser's essential features. This framework allows for an analytically tractable mean-field theory. A more realistic approach would include atom number fluctuations, a Gaussian cavity mode profile, and a distribution of atomic velocities in the $x$ direction. We have numerically confirmed that these effects do not significantly modify the minimum linewidth, maximum power, and minimum pulling coefficient in the superradiant regime.

To realize a superradiant beam laser, one must choose the beam flux, effusive oven design, and cavity parameters to ensure $\Phi \tau^2 \Gamma_c > 8$ and $\delta_D\tau < \pi$. For very narrow linewidths, it may be necessary to reduce $\delta_D$ by transverse laser cooling the atomic beam. Furthermore, to realize a given linewidth, cavity pulling must be kept small enough to prevent excessive broadening from environmental noise. If cavity pulling remains minuscule, the linewidth can be narrowed by decreasing the cavity finesse and increasing $\Phi$; however, the trade-off is that in the limit of extremely small cavity finesse, the laser power vanishes as atoms radiate appreciably into other modes. We note that if the atomic beam is aggressively cooled such that $\delta_D$ is comparable to the recoil frequency, then optomechanical effects are required to model the beam laser correctly~\cite{Jager2019}.

Superradiant lasers based on cold atoms have achieved impressive results, but parasitic heating from atomic repumping has so far limited these systems to pulsed operation~\cite{norcia2016coldSrLaser}. The beam laser design avoids the heating problem since pumping is performed outside the cavity (Fig.~\ref{fig: system}). Therefore, the beam laser configuration may be a more promising approach for realizing a CW superradiant laser. Furthermore, our design could conceivably be made simpler and less fragile than cold-atom or cavity-stabilized systems. For this reason, the superradiant beam laser may be well suited to operate in accelerating frames, making this design potentially useful for space technology, inertial sensors, geodesy, field-based magnetometry, and astrophysical measurements. We hope that our design will make ultracoherent lasers, which are currently limited to a handful of specialized labs, ubiquitous in quantum science.
  
\begin{acknowledgments}
We would like to thank J.P. Bartolotta, J. Cooper, and J.K. Thompson for useful discussions. This research is supported by the Research Centres of Excellence program supported by the National Research Foundation (NRF) Singapore; the Ministry of Education, Singapore; the NSF AMO Grant No. 1806827; NSF PFC Grant No. 1734006; and the DARPA and ARO Grant No. W911NF-16-1-0576. M.H. acknowledges support from a Visiting Fellowship from the University of Pisa, Italy.
\end{acknowledgments}


\begin{thebibliography}{99}
\bibitem{footnote1}
        {The conditions for neglecting free space spontaneous emission are that $N \Gamma_c \gg \gamma$ and $\gamma \tau \ll 1$. We simulated this system with free space spontaneous emission and confirmed that our model (which neglects this effect) reproduces the correct minimum $\Delta \omega$, maximum $P$, and minimum $\wp$.}
\end{thebibliography}
\end{document}


\title{Supplemental Material: Rugged mHz-Linewidth Superradiant Laser Driven by a Hot Atomic Beam}
\author{Haonan~Liu}
\affiliation{JILA, National Institute of Standards and Technology, and University of Colorado, Boulder, Colorado 80309-0440, USA}
\author{Simon~B.~Jäger}
\affiliation{JILA, National Institute of Standards and Technology, and University of Colorado, Boulder, Colorado 80309-0440, USA}
\author{Xianquan~Yu}
\affiliation{Centre for Quantum Technologies, Department of Physics, National University of Singapore, Singapore 117543}
\author{Steven~Touzard}
\affiliation{Centre for Quantum Technologies, Department of Physics, National University of Singapore, Singapore 117543}
\author{Athreya~Shankar}
\affiliation{JILA, National Institute of Standards and Technology, and University of Colorado, Boulder, Colorado 80309-0440, USA}
\author{Murray~J.~Holland}
\affiliation{JILA, National Institute of Standards and Technology, and University of Colorado, Boulder, Colorado 80309-0440, USA}
\author{Travis~L.~Nicholson}
\affiliation{Centre for Quantum Technologies, Department of Physics, National University of Singapore, Singapore 117543}

\maketitle
\section{Theoretical and Numerical Description of the Beam Laser\label{sec: theoryModel}}
In this section we provide the foundations for the theoretical description of the beam laser system. We will apply this description to derive the $c$-number Langevin equations that are used to simulate the dynamics of the atoms. Then we describe in detail how to numerically find the linewidth $\Delta \omega$, the output power $P$, and the pulling coefficient $\wp$ that are visible in Fig.~2b--c, Fig.~3, and Fig.~4 of the main text. In the end of this section we provide a discussion on free-space spontaneous emission. 

\subsection{Quantum mechanical description of the beam laser\label{S11}}
The coherent dynamics of atoms coupled to a single cavity mode is governed by the Hamiltonian
\begin{equation}
    \label{eqn: Hamiltonian_beamLaser}
    \hat{H} \equiv \hbar \Delta {\hat{a}}^\dag \hat{a} 
            + \frac{\hbar g}{2} \sum_j \eta(\textbf{x}_j) \left(\spj \hat{a} + \hat{a}^\dag \smj\right),
\end{equation}
reported in the interaction picture rotating with the frequency of the atomic transition. Here, ${\hat{a}}^\dag$ and $\hat{a}$ are field creation and annihilation operators and $\hat{\sigma}_j^+ = \left(\hat{\sigma}_j^-\right)^\dagger = \ket{e}_j\bra{g}_j$ are raising and lowering operators of the $j$th atomic dipole with electronic excited state $\ket{e}_j$ and ground state $\ket{g}_j$. The frequency $\Delta = \omega_c - \omega_a$ is the detuning between the resonance frequency $\omega_c$ of the cavity mode and the atomic transition frequency $\omega_a$. We define $g$ to be the vacuum Rabi frequency at a cavity antinode, and $\eta(\mathbf{x})$ to be a cavity mode function evaluated at position $\mathbf{x}$. We ignore the effect of optomechanical forces on the atom during the transit time $\tau$. This is justified if the atomic momentum width exceeds the momentum exchange of the atoms with the cavity field during $\tau$. Consequently, we assume that the atoms move ballistically through the cavity, i.e., 
\begin{equation}
    \label{eqn: ballisticMotion}
    \difft{\textbf{x}_j} = \frac{\textbf{p}_j}{m},
\end{equation}
where $m$ is the atomic mass, $\textbf{x}_j(t)$ is the position of atom $j$, and $\textbf{p}_j$ is the constant momentum of atom $j$.

Besides the coherent effects established by the Hamiltonian in Eq.~\eqref{eqn: Hamiltonian_beamLaser}, we also include cavity photon losses with rate $\kappa$~\cite{footnote1}. These give rise to dissipation and noise in the coupled atom-cavity dynamics. These effects are described by quantum Langevin equations (QLEs)~\cite{gardiner1985input} that read
\begin{align}
    \label{eqn: QLEs_field_a}
    \difft{\hat{a}} &= - \left(\frac{\kappa}{2} + i \Delta \right) \hat{a} - \frac{i g}{2} \hat{J}^{-} - \sqrt{\kappa} \hat{\xi},\\
    \difft{\hat{a}^\dag} &= - \left(\frac{\kappa}{2} - i \Delta \right) \hat{a}^\dag + \frac{i g}{2} \hat{J}^{+} - \sqrt{\kappa} \hat{\xi}^\dag,\\
    \label{eqn: QLEs_field_smj}
    \difft{\smj} &= \frac{i g}{2} \eta_j \szj \hat{a}, \\
    \label{eqn: QLEs_field_spj}
    \difft{\spj} &= - \frac{i g}{2} \eta_j \hat{a}^\dag \szj, \\
    \label{eqn: QLEs_field_szj}
    \difft{\szj} &= i g \eta_j \left(\hat{a}^{\dag} \smj - \spj \hat{a} \right),
\end{align}
where we have introduced the collective dipole
\begin{equation}
    \hat{J}^{\pm} \equiv \sum_{j} \eta_{j} \hat{\sigma}^{\pm}_j,\label{quantumcollectivedipole}
\end{equation}
and $\eta_{j} \equiv \eta(\textbf{x}_j)$.
Furthermore we have defined $\szj = \spj\smj - \smj\spj$ and the noise operators $\hat{\xi}$ and $\hat{\xi}^\dag$ that satisfy the correlations $\langle\hat{\xi}(t)\hat{\xi}^\dag(t')\rangle = \delta(t-t')$ and $\langle\hat{\xi}(t)\hat{\xi}(t')\rangle = \langle\hat{\xi}^\dag(t)\hat{\xi}^\dag(t')\rangle = \langle\hat{\xi}^\dag(t)\hat{\xi}(t')\rangle = 0$.

As we refer to in the main text, we analyze the coupled dynamics between the atoms and cavity in the bad cavity limit where $\kappa \gg \{1/\tau, \sqrt{N}g, \delta_D\}$. Here, $N$ is the intracavity atom number and $\delta_D$ is the Doppler width of the atomic beam in the direction of the cavity axis. In this limit the cavity degrees of freedom evolve much faster than the atomic degrees of freedom and can therefore be adiabatically eliminated. This elimination relies on a coarse graining $\Delta t$ in time, where $\Delta t$ is much longer than the cavity relaxation time but much shorter than the typical relaxation time of the atomic degrees of freedom. On this timescale, the field operator $\hat{a}$ can be approximated by
\begin{align}
    \label{eqn: adiabaticElimination}
    \hat{a}\approx -\frac{\left(\gd + i \gc \right)}{g}\hat{J}^{-} - \frac{\left(\gc - i \gd \right)}{g} \frac{2}{\sqrt{\gzero}}\hat{\xi}_{\mathrm{eff}},
\end{align}
where
\begin{equation}
    \gc \equiv \frac{g^{2}\kappa/4}{\kappa^{2}/4 + \Delta^{2}},\label{eqn: Gammac}\ \ \ \ \ \ \gd \equiv \frac{g^{2}\Delta/2}{\kappa^{2}/4 + \Delta^{2}}, \ \ \ \ \ \ \gzero \equiv \frac{g^{2}}{\kappa}. 
\end{equation}
We have introduced effective noise variables $\hat{\xi}_{\mathrm{eff}}$ and $\hat{\xi}_{\mathrm{eff}}^\dag$ that are $\delta$-correlated on the coarse-grained timescale and fulfill $\langle\hat{\xi}_{\mathrm{eff}}(t)\hat{\xi}_{\mathrm{eff}}^\dag(t')\rangle = \delta(t-t')$ and $\langle\hat{\xi}_{\mathrm{eff}}(t)\hat{\xi}_{\mathrm{eff}}(t')\rangle = \langle\hat{\xi}_{\mathrm{eff}}^\dag(t)\hat{\xi}_{\mathrm{eff}}^\dag(t')\rangle = \langle\hat{\xi}_{\mathrm{eff}}^\dag(t)\hat{\xi}_{\mathrm{eff}}(t')\rangle = 0.$

\vspace{1mm}

Using Eq.~\eqref{eqn: adiabaticElimination}, the resulting dynamics of the atomic variables are described by the equations
\begin{align}
    \label{eqn: QLEs_AE_noGamma_noT2_smj}
    \difft{\smj} = &\frac{1}{2} \left(\gc - i \gd \right) \eta_{j} \szj \hat{J}^{-}
                    - \left(\gd +  i \gc \right) \frac{\eta_{j}}{\sqrt{\gzero}} \szj \hat{\xi}_{\mathrm{eff}}, \\
    \label{eqn: QLEs_AE_noGamma_noT2_spj}
    \difft{\spj} = &\frac{1}{2} \left(\gc + i \gd \right) \eta_{j} \hat{J}^{+} \szj 
                    - \left(\gd -  i \gc \right) \frac{\eta_{j}}{\sqrt{\gzero}} \hat{\xi}_{\mathrm{eff}}^{\dag}  \szj, \\                    
    \label{eqn: QLEs_AE_noGamma_noT2_szj}
    \difft{\szj} = &- \left(\gc + i \gd \right) \eta_{j} \hat{J}^{+} \smj
                    - \left(\gc - i \gd \right) \eta_{j} \spj \hat{J}^{-} \notag\\
                   &+ \left(\gd - i \gc \right)  \frac{2 \eta_{j}}{\sqrt{\gzero}} \hat{\xi}_{\mathrm{eff}}^{\dag} \smj
                    + \left(\gd + i \gc \right)  \frac{2 \eta_{j}}{\sqrt{\gzero}} \spj \hat{\xi}_{\mathrm{eff}}.
\end{align}

\subsection{Derivation and simulation of the $c$-number Langevin equations\label{S12}}
Equations~\eqref{eqn: QLEs_AE_noGamma_noT2_smj}--\eqref{eqn: QLEs_AE_noGamma_noT2_szj} describe the full quantum dynamics of the atomic variables when the cavity degrees can be eliminated. However, these equations are in general impossible to solve and hence we derive a semiclassical description of the atomic variables that allows the system to be tractable. This semiclassical description relies on the simulation of $c$-numbers for the dipole components of the atoms traversing the cavity. 

In order to derive the $c$-number semiclassical equations, we first introduce the Hermitian operators
\begin{equation}
    \label{eqn: pm-xy}
    \twovec{\hat{\sigma}^{x}}{\hat{\sigma}^{y}} \equiv R \twovec{\hat{\sigma}^{+}}{\hat{\sigma}^{-}},\ \ \ 
    \twovec{\hat{\xi}^{q}}{\hat{\xi}^{p}} \equiv R \twovec{\hat{\xi}_{\mathrm{eff}}^{\dag}}{\hat{\xi}_{\mathrm{eff}}},
\end{equation}
where $R = \twomatrix{1}{1}{-i}{i}$. These Hermitian operators are observables, and in particular $(\hat{\sigma}^x_j,\hat{\sigma}^y_j,\hat{\sigma}^z_j)$ is the quantum dipole of the atom indexed by $j$. Using the definitions in Eq.~\eqref{eqn: pm-xy} we rewrite Eqs.~\eqref{eqn: QLEs_AE_noGamma_noT2_smj}--\eqref{eqn: QLEs_AE_noGamma_noT2_szj} and obtain
\begin{align}
    \label{eqn: QLEs_AE_noGamma_noT2_sxj}
    \difft{\sxj}  = &\frac{\gc}{2} \eta_{j} \left(\sum_{k \neq j}\eta_{k}\sxk\szj - \eta_{j}\sxj\right)
                   - \frac{\gd}{2} \eta_{j} \left(\sum_{k \neq j}\eta_{k}\syk\szj - \eta_{j}\syj\right) 
                   - \frac{\eta_{j}}{\sqrt{\gzero}} \szj \left(\gc\hat{\xi}^{p} + \gd\hat{\xi}^{q}\right), \\
    \label{eqn: QLEs_AE_noGamma_noT2_syj}
    \difft{\syj} &=  \frac{\gc}{2} \eta_{j} \left(\sum_{k \neq j}\eta_{k}\syk\szj - \eta_{j}\syj\right)
                   + \frac{\gd}{2} \eta_{j} \left(\sum_{k \neq j}\eta_{k}\sxk\szj - \eta_{j}\sxj\right) 
                   + \frac{\eta_{j}}{\sqrt{\gzero}} \szj \left(\gc\hat{\xi}^{q} - \gd\hat{\xi}^{p}\right), \\
    \label{eqn: QLEs_AE_noGamma_noT2_szj_xyz}
    \difft{\szj} &= - \gc \eta_{j}^2\left(\szj + 1\right)
                    - \frac{\gc}{2} \eta_{j} \sum_{k \neq j}\eta_{k}\left(\sxk\sxj + \syk\syj\right) 
                    + \frac{\gd}{2} \eta_{j} \sum_{k \neq j}\eta_{k}\left(\syk\sxj - \sxk\syj\right) \notag\\
                   &+ \frac{\eta_{j}}{\sqrt{\gzero}} \left(\gc\left(\sxj\hat{\xi}^{p} - \syj\hat{\xi}^{q}\right) 
                                                         + \gd\left(\sxj\hat{\xi}^{q} + \syj\hat{\xi}^{p}\right)\right).
\end{align}

We now perform the $c$-number approximation for the Hermitian quantum operators~\cite{schachenmayer2015monteCarlo} that relies on the following replacements $\hat{\sigma}^{\mu}_j \rightarrow s^\mu_j,\,\mu \in \{x, y, z\}$ and $\hat{\xi}^\nu \rightarrow \xi^\nu,\,\nu \in \{q, p\}$, where $s^\mu_j$ is the classical dipole component of atom $j$ and $\xi^\nu$ are classical noise processes with $\langle\xi^\nu(t)\xi^\mu(t')\rangle=\delta_{\mu\nu}\delta(t-t')$. The $c$-number equations are derived by first performing a symmetric ordering of the operators in Eqs.~\eqref{eqn: QLEs_AE_noGamma_noT2_sxj}--\eqref{eqn: QLEs_AE_noGamma_noT2_szj_xyz} and then exchanging the quantum variables by their corresponding $c$-number~\cite{tieri2015thesis} equivalents. This leads to the $c$-number Langevin equations 
\begin{align}
    \label{eqn: cQLEs_AE_noGamma_noT2_sxj_xyz}
    \difft{s_j^x} &= \frac{\gc}{2}\eta_j\left(\mj^x s_j^z - \eta_j s_j^x (s_j^z + 1)\right) 
                    - \frac{\gd}{2}\eta_j\left(\mj^y s_j^z - \eta_j s_j^y (s_j^z + 1)\right) - \frac{\gc}{\sqrt{\gzero}}\eta_j s_j^z \xi^p - \frac{\gd}{\sqrt{\gzero}}\eta_j s_j^z \xi^q, \\
    \label{eqn: cQLEs_AE_noGamma_noT2_syj_xyz}                
    \difft{s_j^y} &= \frac{\gc}{2}\eta_j\left(\mj^y s_j^z - \eta_j s_j^y (s_j^z + 1)\right)
                    + \frac{\gd}{2}\eta_j\left(\mj^x s_j^z - \eta_j s_j^x (s_j^z + 1)\right) + \frac{\gc}{\sqrt{\gzero}}\eta_j s_j^z \xi^q - \frac{\gd}{\sqrt{\gzero}}\eta_j s_j^z \xi^p,\\
    \label{eqn: cQLEs_AE_noGamma_noT2_szj_xyz}                
    \difft{s_j^z} &= - \gc \eta_j^2 (s_j^z + 1) 
                      - \frac{\gc}{2}\eta_j\left(\mj^x s_j^x + \mj^y s_j^y - \eta_j\left(\left(s_j^x\right)^2 + \left(s_j^y\right)^2\right)\right)
                      + \frac{\gd}{2}\eta_j\left(\mj^y s_j^x - \mj^x s_j^y\right) \notag\\
                      &+ \frac{\gc}{\sqrt{\gzero}}\eta_j \left(s_j^x \xi^p - s_j^y \xi^q\right) + \frac{\gd}{\sqrt{\gzero}}\eta_j \left(s_j^x \xi^q + s_j^y \xi^p\right),
\end{align}
where
\begin{equation}
    \mj^x = \sum_j \eta_j s_j^x, \ \ \ \ \mj^y = \sum_j \eta_j s_j^y
\end{equation}
are $c$-number values for the collective dipole components.

For the results that are shown in the main text we have numerically integrated Eqs.~\eqref{eqn: cQLEs_AE_noGamma_noT2_sxj_xyz}--\eqref{eqn: cQLEs_AE_noGamma_noT2_szj_xyz}. Specifically, for the resonant case $\Delta = 0$, we obtain Eqs.~(1)--(3) of the main text. 

When running the $c$-number simulations, we use a Monte Carlo method by initializing and running many trajectories of the time series of position variables $\textbf{x}_j$ and dipole variables $s^\mu_j,\,\mu \in \{x, y, z\}$. We find that the simulations converge with low error when the number of trajectories is $\gtrsim 50$. We define the phase space as ${\bf x}=(x,z)$ and ${\bf p}=(p_x,p_z)$, with the $z$-direction aligned along the cavity axis and the $x$-direction aligned along the direction of the atomic beam (see Fig.~1 in the main text). For simplicity, we consider an explicit form of $\eta({\bf x})$ 
\begin{equation}
    \eta({\bf x})=\cos(k z)\left[\Theta(x+w)-\Theta(x-w)\right],\label{eta0}
\end{equation}
where $w$ is the cavity beam waist, $k=2\pi/\lambda$ is the wavenumber, and $\lambda$ is the wavelength of the cavity mode. In each trajectory, for atom $j$ that enters the cavity, we initialize the positions by taking $x_j = -w$, and sampling $z_j$ from a uniform distribution with a range of $\lambda$. Here we impose a periodic boundary condition in the $z$-direction, which is valid when the radius of the atomic beam is much larger than $\lambda$. We initialize the momentum by taking $p_{x,j}/m \equiv 2w/\tau$, and randomly sampling $p_{z,j}$ from a Maxwell-Boltzmann distribution at a given temperature. Since atoms undergo ballistic motion as shown in Eq.~\eqref{eqn: ballisticMotion}, $p_{x,j}$ and $p_{z,j}$ are constant for each trajectory. At each timestep $\delta t = \tau/N$, precisely one atom enters the cavity, and any atom that has accumulated a time $\tau$ inside the cavity leaves. Consequently, after $\tau$, the intracavity atom number becomes stationary, $i.e.$ $N\equiv\Phi\tau$.

The dipole variables are initialized in the following way~\cite{schachenmayer2015monteCarlo}: in each trajectory, for atom $j$ entering the cavity, we randomly choose $s_j^x$ and $s_j^y$ to be $\pm1$ with equal probability, and $s_j^z$ as 1. This ensures that the correct second moments of the $c$-number variables are retrieved according to those of the symmetric ordered second moments of the quantum operators, $i.e.$ $\langle s^\mu_j \rangle = \langle \hat{\sigma}^{\mu}_j \rangle$, $\langle s_j^xs_j^y \rangle = \langle \{\hat{\sigma}_j^x\hat{\sigma}_j^y\}_{\mathrm{sym}} \rangle = 0$, $\langle (s_i^x)^2 \rangle = \langle (s_i^y)^2 \rangle = \langle (\hat{\sigma}_i^x)^2 \rangle = \langle (\hat{\sigma}_i^y)^2 \rangle = 1$, etc. Here, the average $\langle\,\cdot \,\rangle$ applied to $c$-numbers represents the trajectory average over different possible initializations and sampled values of the the stochastic noises $\xi^q$ and $\xi^p$, and $\{\cdot\}_{\mathrm{sym}}$ is the symmetric ordering of the operators inside.

\subsection{Numerical calculation of the linewidth, the output power, and the pulling coefficient}

We will now explain how we numerically calculated the linewidth $\Delta \omega$ in Fig.~2b--c, the laser output power $P$ in Fig.~3, and the pulling coefficient $\wp$ in Fig.~4 from $c$-number simulations.
The linewidths shown in Fig.~2b--c of the main text were extracted from fits to the real part of the first-order two-time correlation function $g^{(1)}$ defined as
\begin{align}
    g^{(1)}(t,t')=\left\langle \hat{a}^{\dag}(t+t')\hat{a}(t')\right\rangle\propto\left\langle \hat{J}^{+}(t+t')\hat{J}^{-}(t')\right\rangle,
\end{align}
where the cavity field is adiabatically eliminated as in Eq.~\eqref{eqn: adiabaticElimination} and $t'$ is some time after the system reaches steady state. For the actual calculation we first use our $c$-number simulations to derive 
\begin{align}
    \mathrm{Re}[g^{(1)}(t,t')] \propto \left\langle\mj^{x}(t + t')\mj^{x}(t')\right\rangle + \left\langle\mj^{y}(t + t')\mj^{y}(t')\right\rangle,
\end{align} 
After that we average over different times $t'$ to obtain $\bar{g}^{(1)}(t) = \left\langle\mathrm{Re}[g^{(1)}(t,t')]\right\rangle_{t'}$, where $\langle\,.\,\rangle_{t'}$ the time average. We fit the result $\bar{g}^{(1)}(t)$ to a function of form $ F(t) = C_1 e^{-C_2t}\cos{C_3 t}$ and thereby derive fitting parameters $C_1$, $C_2$, and $C_3$. The linewidth of the light field is \begin{align}
    \Delta\omega=2C_2.
\end{align}
For the resonant case, $\Delta=0$, shown in Fig.~2c, we observe $C_3 = 0$ for all data points except when $\Phi\tau^2\Gamma_c \gtrsim 40$ for the $\delta_D\tau = 2\pi$ curve. For these parameters we observe oscillations in $\bar{g}^{(1)}$ resulting in a non-vanishing $C_3$. Here we anticipate that the system will be in an unstable region of parameter space.

For the non-resonant case, $\Delta\neq0$, with $\delta_D\tau = 0.2\pi$, we also observe oscillations in $\bar{g}^{(1)}$ and therefore obtain $C_3\neq0$. The reason for this is that the frequency of the output field $\omega$ is slightly pulled towards $\omega_c$ and therefore not in perfect resonance with $\omega_a$. By appropriately scaling $C_3$, this allows us to extract the pulling coefficient shown in Fig.~4 of the main text by means of
\begin{align}
\wp=\frac{C_3}{\Delta}.  
\end{align}

The values of output power shown in Fig.~3 of the main text were calculated from \begin{align}
P = \hbar \omega \kappa\int _{t_0}^{T}\,dt\frac{\langle\hat{a}^{\dag}(t)\hat{a}(t)\rangle}{T-t_0}\approx\frac{\hbar \omega\gc}{4} \int _{t_0}^{T}\,dt\frac{\left\langle \left(\mj^{x}(t)\right)^2 + \left(\mj^{y}(t)\right)^2\right\rangle}{T-t_0},
\end{align}
where we performed a time average over the interval $[t_0,T]$. Here, $t_0$ is a time after which the system has reached steady state and $T$ is the total simulation time of the $c$-number Langevin equations.

\subsection{Comments on free-space spontaneous emission}
Free-space spontaneous emission has been neglected in our treatment because our system works in the parameter regime where collective emission into the cavity mode dominates over spontaneous emission into free space (i.e., $N \Gamma_c \gg \gamma$, where $\gamma$ is the spontaneous emission rate)~\cite{meiser2009prospects, meiser2010ssLaser, xu2014synch}. An additional requirement for superradiance is that atomic excitations do not dissipate into the free-space modes external to the cavity during the transit of an atom. This is true if the free-space spontaneous lifetime $1/\gamma$ is large compared to the transit time $\tau$ (i.e., $\gamma\tau\ll1$). In this regime, the atomic dipoles synchronize, and this overcomes broadening from residual free space spontaneous emission.
	
In the following we show the quantum Langevin equations and the corresponding $c$-number Langevin equations that include the description of spontaneous emission. In order to describe this effect in the system, we replace Eqs.~\eqref{eqn: QLEs_field_smj}--\eqref{eqn: QLEs_field_szj} by 
\begin{align}
	\difft{\smj} &= \frac{i g}{2} \eta_j \szj \hat{a} - \frac{\gamma}{2} \smj
	+ \sqrt{\gamma} \szj \hat{\xi}_{j}^{\gamma},\\
	\difft{\spj} &= - \frac{i g}{2} \eta_j \hat{a}^{\dagger} \szj - \frac{\gamma}{2} \spj
	+ \sqrt{\gamma} \szj \left[\hat{\xi}_{j}^{\gamma}\right]^{\dag},\\
	\difft{\szj} &= i g \eta_j \left(\hat{a}^{\dag} \smj - \spj \hat{a} \right) - \gamma \left(\szj + 1\right)
	- 2 \sqrt{\gamma} \left(\spj \hat{\xi}_{j}^{\gamma} + \smj \left[\hat{\xi}_{j}^{\gamma}\right]^{\dag}\right),
\end{align}
where $\hat{\xi}_{j}^{\gamma}$ and $\left[\hat{\xi}_{j}^{\gamma}\right]^{\dag}$ are the noise terms that arise from spontaneous emission. They have vanishing first moments and satisfy $\Big\langle\hat{\xi}_j^{\gamma}(t)\left[\hat{\xi}_k^{\gamma}\right]^\dag(t')\Big\rangle = \delta_{jk}\delta(t-t')$ and $\Big\langle\hat{\xi}_j^{\gamma}(t)\hat{\xi}_k^{\gamma}(t')\Big\rangle = 0=  \Big\langle\left[\hat{\xi}_j^{\gamma}\right]^\dag(t)\hat{\xi}_k^{\gamma}(t')\Big\rangle$. 
	
To solve these equations, we eliminate the cavity degrees of freedom in the regime where the lifetime of the cavity photons is much shorter than any typical atomic relaxation timescale, i.e. $\kappa \gg \{\gamma, 1/\tau, \sqrt{N}g, \delta_D\}$. This is done as described in Sec.~S~1.1. Next, we use a semiclassical description to simulate the quantum Langevin equations. The semiclassical model is derived using the same steps as in Sec.~S~1.2 and uses $c$-number Langevin equations to simulate the dynamics of the dipole components. 
	
These equations replace Eqs.~\eqref{eqn: cQLEs_AE_noGamma_noT2_sxj_xyz}--\eqref{eqn: cQLEs_AE_noGamma_noT2_szj_xyz} and are given by
\begin{align}
	\difft{s_j^x} &= \frac{\gc}{2}\eta_j\left(\mj^x s_j^z - \eta_j s_j^x (s_j^z + 1)\right) 
	- \frac{\gd}{2}\eta_j\left(\mj^y s_j^z - \eta_j s_j^y (s_j^z + 1)\right) 
	- \frac{\gamma}{2} s_j^x  \notag\\ 
	&- \frac{\gc}{\sqrt{\gzero}}\eta_j s_j^z \xi^p 
	- \frac{\gd}{\sqrt{\gzero}}\eta_j s_j^z \xi^q 
	+ \mathcal{F}_{j}^{x},\label{gammasx}\\
	\difft{s_j^y} &= \frac{\gc}{2}\eta_j\left(\mj^y s_j^z - \eta_j s_j^y (s_j^z + 1)\right)
	+ \frac{\gd}{2}\eta_j\left(\mj^x s_j^z - \eta_j s_j^x (s_j^z + 1)\right) 
	- \frac{\gamma}{2} s_j^y \notag\\
	&+ \frac{\gc}{\sqrt{\gzero}}\eta_j s_j^z \xi^q 
	- \frac{\gd}{\sqrt{\gzero}}\eta_j s_j^z \xi^p
	+ \mathcal{F}_{j}^{y},\label{gammasy}\\
	\difft{s_j^z} &= - \left(\gc \eta_j^2 + \gamma \right) (s_j^z + 1) 
	- \frac{\gc}{2}\eta_j\left(\mj^x s_j^x + \mj^y s_j^y - \eta_j\left(\left(s_j^x\right)^2 + \left(s_j^y\right)^2\right)\right)
	+ \frac{\gd}{2}\eta_j\left(\mj^y s_j^x - \mj^x s_j^y\right) \notag\\
	&+ \frac{\gc}{\sqrt{\gzero}}\eta_j \left(s_j^x \xi^p - s_j^y \xi^q\right) 
	+ \frac{\gd}{\sqrt{\gzero}}\eta_j \left(s_j^x \xi^q + s_j^y \xi^p\right)
	+ \mathcal{F}_{j}^{z}.\label{gammasz}
\end{align}
Beside the terms that have already been described within Eqs.~\eqref{eqn: cQLEs_AE_noGamma_noT2_sxj_xyz}--\eqref{eqn: cQLEs_AE_noGamma_noT2_szj_xyz}, the new equations also simulate spontaneous emission. This mechanism gives rise to a decay of coherence (shown in Eqs.~\eqref{gammasx}--\eqref{gammasy}) and it de-excites the atoms (Eq.~\eqref{gammasz}). In addition, we simulate noise terms $\mathcal{F}_{j}^{a}$ with $a\in\{x,y,z\}$ that arise from spontaneous emission. These terms have zero mean, and their second moments are described by the correlation
\begin{align}
\label{eqn: diffusion}
	\left\langle \mathcal{F}_{j}^{a}(t)\mathcal{F}_{k}^{b}(t')\right\rangle=2D^{ab}_j\delta_{jk}\delta(t-t'),
\end{align}
where the symmetric diffusion matrix is
\begin{align}
		D^{ab}_j \equiv\begin{pmatrix}
			D^{xx}_j&D^{xy}_j&D^{xz}_j\\
			D^{yx}_j&D^{yy}_j&D^{yz}_j\\
			D^{zx}_j&D^{zy}_j&D^{zz}_j\\
		\end{pmatrix}
		=\frac{\gamma}{2}\begin{pmatrix}
			1 & 0 & s^x_j \\
			0 & 1 & s^y_j \\
			s^x_j &  s^y_j & 2\left(1+ s^z_j\right)\\
		\end{pmatrix},\ \ \ a,b\in\{x,y,z\}.
\end{align} 
Here the $\langle\cdot\rangle$ used in Eq.~\eqref{eqn: diffusion} means the expectation value over the reservoir variables. 

We have simulated Eqs.~\eqref{gammasx}--\eqref{gammasz} in the regime where $\gamma\ll \{N\Gamma_c,~\tau^{-1}\}$ and confirmed that changes in the minimum linewidth, maximum power, and minimum pulling coefficient are negligible. 
\section{Mean-field analysis}
In this section we derive and analyze a mean-field description of Eqs.~\eqref{eqn: cQLEs_AE_noGamma_noT2_sxj_xyz}--\eqref{eqn: cQLEs_AE_noGamma_noT2_szj_xyz}. Our mean field model uses a classical phase-space density for the dipoles
\begin{equation}
    s_{\mu}({\bf x},{\bf p},t) = \sum_{j} s_j^{\mu} \delta({\bf x}-{\bf x}_j)\delta({\bf p}-{\bf p}_j),\ \ \ \ \mu \in \{ x, y, z\},
\end{equation}
where we use ${\bf x}=(x,z)$ and ${\bf p}=(p_x,p_z)$.

In the limit $N \gg 1$, we derive the equations of motion for the average densities $\langle s_x\rangle$, where we factorize second moments and discard noise contributions. Using Eq.~\eqref{eqn: ballisticMotion} and Eqs.~\eqref{eqn: cQLEs_AE_noGamma_noT2_sxj_xyz}--\eqref{eqn: cQLEs_AE_noGamma_noT2_szj_xyz} for the resonant case $\Delta=0$, we find
\begin{align}
    \label{eqn: meanfield sx}
    \frac{\partial \langle s_x\rangle}{\partial t}+\frac{\bf p}{m}\cdot\nabla_{\bf x}\langle s_{x}\rangle=&\frac{\Gamma_c}{2}\eta({\bf x})\langle J_x\rangle \langle s_z\rangle, \\
    \label{eqn: meanfield sy}
    \frac{\partial \langle s_y\rangle }{\partial t}+\frac{\bf p}{m}\cdot\nabla_{\bf x}\langle s_{y}\rangle =&\frac{\Gamma_c}{2}\eta({\bf x})\langle J_y\rangle \langle s_z\rangle, \\
    \label{eqn: meanfield sz}
    \frac{\partial \langle s_z\rangle}{\partial t}+\frac{\bf p}{m}\cdot\nabla_{\bf x}\langle s_{z}\rangle =&-\frac{\Gamma_c}{2}\eta({\bf x})\left[\langle J_x\rangle \langle s_x\rangle+\langle J_y\rangle \langle s_y\rangle\right],
\end{align}
with the collective dipole components defined as
\begin{equation}
    J_{\mu}(t) = \int d{\bf x}d{\bf p} \eta({\bf x}) s_{\mu}({\bf x},{\bf p},t),\ \ \ \ \mu \in \{ x, y\}.
\end{equation}
Here, we have introduced the gradient $\nabla_{\bf x}=(\partial_x,\partial_z)$ and the integration in phase space is defined as $\int d{\bf x}d{\bf p}f({\bf x},{\bf p})=\int_{-w}^wdx\int_{0}^{\lambda} dz\int_{0}^{\infty}dp_x\int_{-\infty}^{\infty}dp_zf(x,z,p_x,p_z)$ for any function $f({\bf x},{\bf p})=f(x,z,p_x,p_z)$.

\vspace{0.3cm}

With $\eta({\bf x})$ defined in Eq.~\eqref{eta0}, the boundary conditions for Eqs.~\eqref{eqn: meanfield sx}--\eqref{eqn: meanfield sz} are
\begin{align}
    \langle s_x(x=-w,z,p_x,p_z)\rangle=&0,\\
    \langle s_y(x=-w,z,p_x,p_z)\rangle=&0,\\
    \langle s_z(x=-w,z,p_x,p_z)\rangle=&\rho({\bf p}),
\end{align}
for the density of atoms that enter the cavity. We consider the case of a spatially homogeneous phase-space density of atoms entering the cavity given by
\begin{align}
    \label{eqn: Density}
    \rho({\bf p})=\frac{N}{2w\lambda}\sqrt{\frac{\beta}{2m\pi}}e^{-\beta\frac{p_z^2}{2m}}\delta\left(p_x-m\frac{2w}{\tau}\right).
\end{align}

This density assumes no broadening in the velocity $v_x=p_x/m$ perpendicular to the cavity axis. Furthermore, we have introduced the inverse temperature $\beta$ in the direction of the cavity axis that determines the Doppler width $\delta_{D}=k/\sqrt{m\beta}$. The density is normalized according to the number of atoms inside of the cavity
\begin{align}
N=\int d{\bf x}d{\bf p}\rho({\bf p}).
\end{align}

\subsection{The non-superradiant regime}

In this subsection we investigate the non-superradiant atomic beam configuration and its stability. This stability analysis is used to derive the mean-field linewidth that is visible in Fig.~2a--b of the main text. The non-superradiant phase is a stationary solution of Eqs.~\eqref{eqn: meanfield sx}--\eqref{eqn: meanfield sz} given by ${\langle s_x\rangle=0=\langle s_y\rangle}$, implying a vanishing mean-field value of the collective dipoles ${\langle J_x\rangle=0=\langle J_y\rangle}$. In this non-superradiant regime the atoms travel through the cavity and remain in the excited state  
\begin{align}
\langle s_{z}(x,z,p_x,p_z)\rangle=&\rho({\bf p}).
\end{align}

Although this is a stationary solution it is in general not stable. In order to determine the stability of this state we have to derive a linear response theory for small fluctuations $\delta s_m=s_m-\langle s_m\rangle$ that relies on dropping second order terms in $\delta s_m$. In this case the dynamics of $\delta s_x$ and $\delta s_y$ are equivalent and decoupled and hence we can focus on the dynamics of $\delta s_x$. The linearized equation for $\delta s_x$ is given by
\begin{align}
\frac{\partial \delta s_x}{\partial t}+\frac{\bf p}{m}\cdot\nabla_{\bf x}\delta s_x=&\frac{\Gamma_c}{2}\eta({\bf x})\rho({\bf p})\delta J_x,
\end{align}
where $\delta J_x(t)=\int d{\bf x}d{\bf p}\,\eta({\bf x})\delta s_x({\bf x},{\bf p},t)$.
Using the Laplace transform
\begin{align}
L[f](\nu)=\int_{0}^\infty e^{-\nu t}f(t)dt,
\end{align}
we get
\begin{align}
\nu L[\delta s_x]+\frac{\bf p}{m}\cdot\nabla_{\bf x}L[\delta s_x]=\delta s_x(0)+\frac{\Gamma_c}{2}\eta({\bf x})\rho({\bf p})L[\delta J_x],
\end{align}
where $\delta s_x(0)=\delta s_x({\bf x},{\bf p},0)$ is an initial condition.
Solving this equation formally for $L[\delta s_x]$, then multiplying by $\eta({\bf x})$ and integrating over phase space we find 
\begin{align}
L[\delta J_x]=\frac{\int d{\bf x}d{\bf p}\,\int_{0}^\infty dt\,e^{-\nu t}\eta({\bf x}+\frac{{\bf p}}{m}t)\delta s_x({\bf x},{\bf p},0)}{1-\frac{\Gamma_c}{2}\int d{\bf x}d{\bf p}\,\int_{0}^\infty dt\,e^{-\nu t}\eta({\bf x}+\frac{{\bf p}}{m}t)\eta({\bf x})\rho({\bf p})}.
\end{align}
The long-time behavior of $\delta J_x(t)\propto e^{\nu_0 t}$ is governed by the exponent $\nu_0$ that is the zero of the dispersion function
\begin{align}
D(\nu)=&1-\frac{\Gamma_c}{2}\int d{\bf x}d{\bf p}\,\int_{0}^\infty dt\,e^{-\nu t}\eta\left({\bf x}+\frac{{\bf p}}{m}t\right)\eta({\bf x})\rho({\bf p}),\label{Dishom}
\end{align}
with the largest real part. If the real part of this exponent is negative, $\mathrm{Re}(\nu_0)<0$, the non-superradiant state is stable and fluctuations decay with the exponent $\nu_0$. In that case it determines the longest relaxation time and the linewidth of the emitted light.

With the choice of Eq.~\eqref{eta0} and Eq.~\eqref{eqn: Density}, we can explicitly derive the following form of the dispersion relation 
\begin{align}
D(\nu)=&1+\frac{\Phi\tau^2\Gamma_c}{4}\frac{1-e^{-\frac{\delta_{D}^2\tau^2+2\nu\tau}{2}}}{\delta_{D}^2\tau^2}-\frac{\Phi\tau^2\Gamma_c}{4}\sqrt{\frac{\pi}{2\delta_{D}^2\tau^2}}e^{\frac{\nu^2}{2\delta_{D}^2}}\left(1+\frac{\nu\tau}{\delta_{D}^2\tau^2}\right)\left[\mathrm{erf}\left(\frac{\nu+\delta_{D}^2\tau}{\sqrt{2\delta_{D}^2}}\right)-\mathrm{erf}\left(\frac{\nu}{\sqrt{2\delta_{D}^2}}\right)\right],\label{Dispersionerf}
\end{align}
where $\mathrm{erf}(\,.\,)$ is the error function \cite{Abramowitz:1968}. 
We find the roots $\nu_0$ of Eq.~\eqref{Dispersionerf} with the largest real part numerically for different values of $\Phi\tau^2\Gamma_c$ and $\delta_{D}\tau$. In Fig.~2a of the main text we show the linewidth $\Delta\omega$ given by $\Delta\omega=\max\{-2\mathrm{Re}(\nu_0),0\}$ and plotted as contour plot. 

If $\mathrm{Re}(\nu_0)$ becomes positive the non-superradiant regime is unstable and $\mathrm{Re}(\nu_0)$ determines the timescale for the build up of macroscopic coherence in the atomic dipoles. Assuming that the system will reach a stationary superradiant configuration we will determine this stationary state in the following subsection.

\subsection{Superradiant configuration}
The purpose of this subsection is the mean-field description of the superradiant phase. With this description we are able to predict a stationary output power of the laser that is visible in Fig.~3 of the main text. In the superradiant phase the system will spontaneously choose a phase of the dipole. We will assume that without loss of generality this phase is zero. Therefore, we assume that we obtain $J_{\mathrm{st}} \equiv \langle J_x\rangle \neq0$ and $\langle J_y\rangle=0$ in the mean-field stationary state. We want to remark that in the mean-field description $\langle J_x\rangle \neq0$ and $\langle J_y\rangle=0$ is stationary on arbitrary timescales, therefore the mean-field treatment predicts a zero linewidth in the superradiant phase.

In order to derive the actual form of the superradiant stationary state we first need to solve 
\begin{align}
\frac{\bf p}{m}\cdot\nabla_{\bf x}\langle s_x\rangle=&\frac{\Gamma_c}{2}\eta({\bf x})\langle s_z\rangle J_{\mathrm{st}},\label{s0eq}\\
\frac{\bf p}{m}\cdot\nabla_{\bf x}\langle s_z\rangle=&-\frac{\Gamma_c}{2}\eta({\bf x})\langle s_x\rangle J_{\mathrm{st}}.\label{z0eq}
\end{align}
The solution of these coupled differential equation can be parameterized by
\begin{align}
\langle s_x\rangle=\rho({\bf p})\sin(K({\bf x},{\bf p})),\label{meanfieldSSRsx}\\
\langle s_z\rangle=\rho({\bf p})\cos(K({\bf x},{\bf p})),
\end{align}
with
\begin{align}
K\left(x-w,z,{\bf p}\right)=&\frac{m\Gamma_cJ_{\mathrm{st}}}{kp_z}\sin\left(\frac{p_z}{2p_x}kx\right)\cos\left(k\left[z-\frac{p_z}{2p_x}x\right]\right)
\end{align}
where we used Eq.~\eqref{eta0} for $-w\leq x\leq w$.

The value of $J_{\mathrm{st}}$ has to be calculated self-consistently. This can be done by multiplying Eq.~\eqref{meanfieldSSRsx} by $\eta({\bf x})$ and integrating over the phase space. The result reads
\begin{align}
J_{\mathrm{st}}&=N\int_{-\infty}^{\infty}du\frac{1}{\sqrt{2\pi\delta_D^2}}e^{-\frac{u^2}{2\delta_D^2}}\frac{1-\mathcal{J}_0\left(\frac{\Gamma_cJ_{\mathrm{st}}\tau}{2}\frac{\sin\left(\frac{u\tau}{2}\right)}{\frac{u\tau}{2}}\right)}{\frac{\Gamma_cJ_{\mathrm{st}}\tau}{2}},\label{J}
\end{align}
where $\mathcal{J}_n$ is the Bessel function of order $n$ \cite{Abramowitz:1968}.

It is straightforward to see that one possible solution of this equation is $J_{\mathrm{st}}=0$. However, we are interested in the non-zero solution of this equation. Defining $j_{\mathrm{st}}=J_{\mathrm{st}}/N$ we obtain from Eq.~\eqref{J} the relation
\begin{align}
j_{\mathrm{st}}&=\int_{-\infty}^{\infty}du\frac{1}{\sqrt{2\pi\delta_D^2}}e^{-\frac{u^2}{2\delta_D^2}}\frac{1-\mathcal{J}_0\left(\frac{\Phi\tau^2\Gamma_cj_{\mathrm{st}}}{2}\frac{\sin\left(\frac{u\tau}{2}\right)}{\frac{u\tau}{2}}\right)}{\frac{\Phi\tau^2\Gamma_cj_{\mathrm{st}}}{2}}.\label{j}
\end{align}
Therefore, keeping $\Phi\tau^2\Gamma_c$ and $\delta_{D}$ constant, a solution $j_{\mathrm{st}}\neq0$ of Eq.~\eqref{j} is independent of $N$ and leads to a superradiant scaling of the power
\begin{align}
P=\hbar \omega \kappa\langle \hat{a}^{\dag}\hat{a}\rangle \approx \frac{\hbar \omega\Gamma_c }{4} N^2 j_{\mathrm{st}}^2,
\end{align}
where $\omega$ is the frequency of the laser light that is here resonant with the atomic transition frequency, $\omega=\omega_a$ .

We use this equation to derive the mean-field prediction of the laser power that is shown in Fig.~3 of the main text as black dashed line. For this we have numerically found the solution of Eq.~\eqref{j} for different values of $\Phi\tau^2\Gamma_c$ and for fixed $\delta_{D}\tau=0.2\pi$.

\subsection{Pulling coefficient}
We will now derive the pulling coefficient from our mean-field equations. The pulling coefficient describes the deviation of the actual laser frequency $\omega$ from the atomic resonance frequency $\omega_a$ in presence of a small but non-vanishing detuning $\Delta$. It is defined as
\begin{align}
    \wp \equiv \frac{\omega -\omega_a}{\Delta}   
\end{align}
and is depicted in Fig.~4 of the main text. In order to find an analytical expression for $\wp$ we need to extend the description of Eqs.~\eqref{eqn: meanfield sx}--\eqref{eqn: meanfield sz} such that they describe the dynamics of the atomic variables in presence of a finite detuning $\Delta$. Using Eq.~\eqref{eqn: ballisticMotion} and Eqs.~\eqref{eqn: cQLEs_AE_noGamma_noT2_sxj_xyz}--\eqref{eqn: cQLEs_AE_noGamma_noT2_szj_xyz} we obtain the following mean-field equations
\begin{align}
    \label{s}
    \frac{\partial \langle s\rangle}{\partial t}+\frac{\bf p}{m}\cdot\nabla_{\bf x}\langle s\rangle &=\xi\frac{\Gamma_c}{2}\eta({\bf x})\langle s_z\rangle \langle J\rangle,\\
    \label{z}
    \frac{\partial \langle s_z\rangle}{\partial t}+\frac{\bf p}{m}\cdot\nabla_{\bf x}\langle s_z\rangle &=-\Gamma_c\eta({\bf x})\left[\xi^*\langle J^*\rangle\langle s\rangle+\xi\langle s^*\rangle \langle J\rangle\right],
\end{align}
where $ s=( s_x+is_y)/2$, $J=\int d{\bf x}d{\bf p} \eta({\bf x}) s$, and $(\,.\,)^*$ denotes complex conjugation. The effect of the finite detuning $\Delta$ is visible here in the imaginary part of
\begin{align}
    \xi \equiv 1-i\frac{\Delta}{\kappa/2},
\end{align}
and the modified emission linewidth $\gc$ defined in Eq.~\eqref{eqn: Gammac}.

A mean-field solution of Eq.~\eqref{s}--\eqref{z} can be parameterized by
\begin{align}
\langle s\rangle=&\frac{\rho({\bf p})}{2}e^{-i\phi({\bf x},{\bf p},t)}\sin\left(K({\bf x},{\bf p},t)\right),\\
\langle s_z\rangle=&\rho({\bf p})\cos\left(K({\bf x},{\bf p},t)\right),
\end{align}
with a space, momentum, and time dependent phase $\phi({\bf x},{\bf p},t)$ and angle $K({\bf x},{\bf p},t)$. In the regime of steady-state superradiant emission we expect that $K({\bf x},{\bf p},t)=K({\bf x},{\bf p})$ is time independent and $\phi({\bf x},{\bf p},t)$ is given by
\begin{align}
\phi({\bf x},{\bf p},t)=(\omega-\omega_a) t+\psi({\bf x},{\bf p}), 
\end{align} 
where $\psi$ is a non-explicitly time dependent phase that the atom acquires when it travels through the cavity. Notice that this also describes the solution given in Eqs.~\eqref{s0eq} and \eqref{z0eq} where we had the special case $\omega=\omega_a$ and $\psi=0$ for the resonant case $\Delta=0$. 

Using Eqs.~\eqref{s} and \eqref{z} together with $J=\int d{\bf x}d{\bf p} \eta({\bf x}) s$ we can find an approximate formula for $\omega$ in the limit where $\Delta/\kappa\ll1$. In this limit the solution for $\omega$ is given by
\begin{align}
\omega-\omega_a \approx&\frac{\Delta}{\kappa/2}\frac{J_{\mathrm{st}}}{\int d{\bf x}\int d{\bf p}\rho({\bf p})\sin\left(K\left({\bf x},{\bf p}\right)\right)\int_{0}^{\infty}dt\eta\left({\bf x}+\frac{\bf p}{m}t\right)}.
\end{align}
The pulling coefficient consequently takes the form
\begin{align}
\wp\approx\frac{1}{\kappa}\frac{2J_{\mathrm{st}}}{\int d{\bf x}\int d{\bf p}\rho({\bf p})\sin\left(K\left({\bf x},{\bf p}\right)\right)\int_{0}^{\infty}dt\eta\left({\bf x}+\frac{\bf p}{m}t\right)}. \label{pulling}
\end{align}
The order of magnitude of the integral in the denominator can be estimated by $\sim N\tau$ in the limit where $\delta_{D}\tau\ll1$. The collective dipole can be estimated as $J_{\mathrm{st}}\sim N$. This leads to a scaling $\wp\propto1/(\kappa\tau)$ showing that pulling is suppressed by a factor $\kappa \tau$ as mentioned in the main text. The numerical value of $\wp$ obtained from Eq.~\eqref{pulling} for different values of $\Phi\tau^2\Gamma_c$ and fixed $\delta_{D}\tau=0.2\pi$ is shown as the purple dashed line in Fig.~4 of the main text. 

\section{Second Order Time Correlation $g^{(2)}(0)$}
\begin{figure}[h!]
    \centering
    \includegraphics[width=0.5\linewidth]{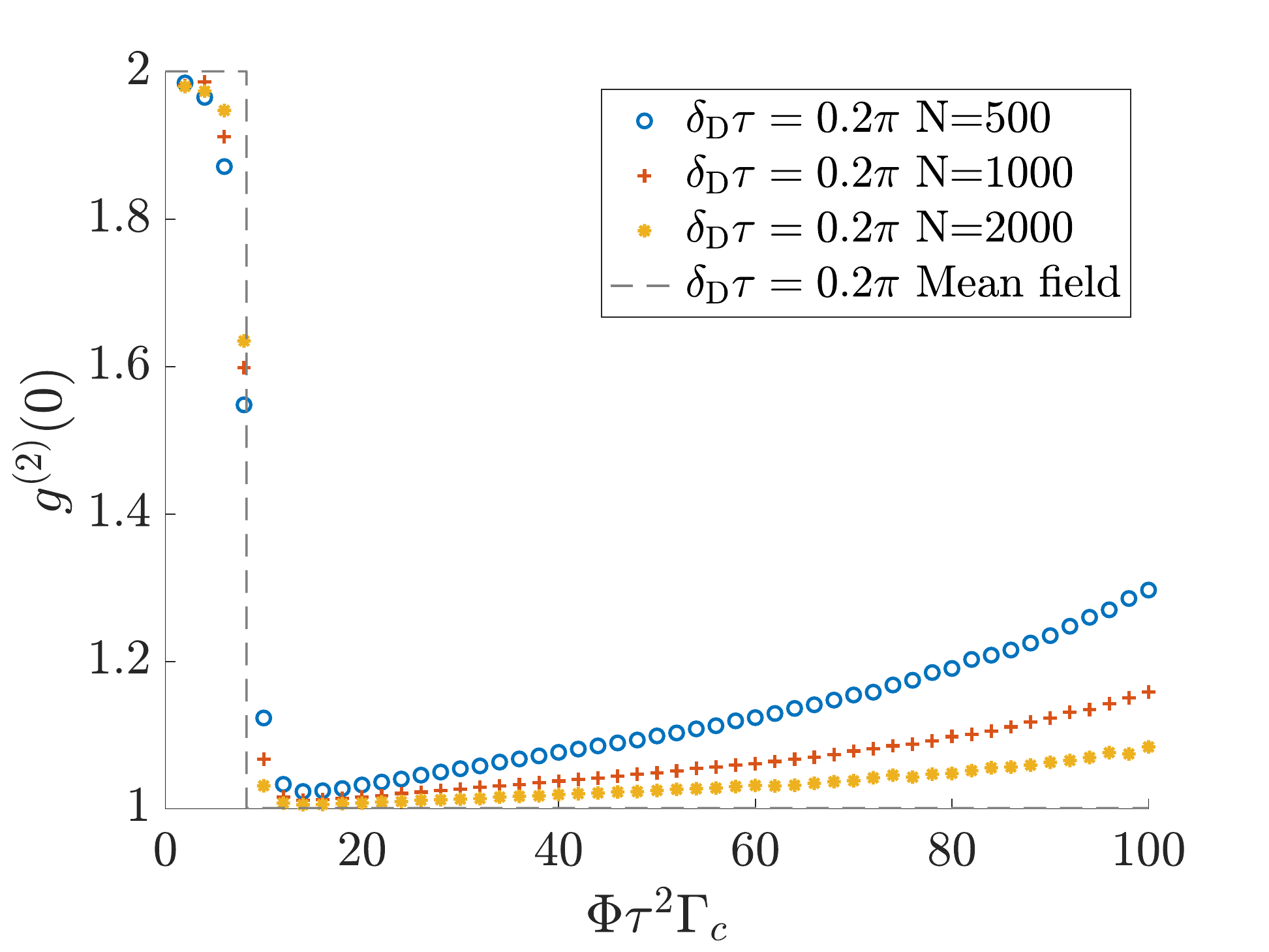}
    \caption{Simulation results of the second-order time correlation function $g^{(2)}(0)$ compared to the mean-field prediction. For each data point, we calculated 100 trajectories each with a simulation time of $T=200$ with $\Phi=500$, $\tau=1$, and $\delta_D\tau=0.2\pi$.}
    \label{fig: g2}
\end{figure}

In this section, we provide supporting evidence that the light field produced by the superradiant beam laser is second order coherent, i.e. $g^{(2)}(0) = 1$. Here, $g^{(2)}(0)$ is defined as
\begin{align}
    g^{(2)}(0)=\frac{\langle \hat{a}^{\dag}\hat{a}^{\dag}\hat{a}\hat{a}\rangle }{\langle \hat{a}^{\dag}\hat{a}\rangle^2}.
\end{align}
The expectation value is here taken at the stationary state. In the limit where the cavity field can be eliminated we can calculate $g^{(2)}(0)$ directly from the collective dipole using
\begin{align}
    g^{(2)}(0)\approx\frac{\langle \hat{J}^{+}\hat{J}^{+}\hat{J}^{-}\hat{J}^{-}\rangle }{\langle \hat{J}^{+}\hat{J}^{-}\rangle^2},
\end{align}
with the definitions in Eq.~\eqref{quantumcollectivedipole}. We use the $c$-number Langevin equations in Eqs.~\eqref{eqn: cQLEs_AE_noGamma_noT2_sxj_xyz}--\eqref{eqn: cQLEs_AE_noGamma_noT2_szj_xyz} with $\delta_D\tau = 0.2\pi$ and $\Delta=0$ to calculate the values of $g^{(2)}(0)$. The results are shown in Fig.~S\ref{fig: g2} for various values of $\Phi\tau^2\Gamma_c$ and for different intracavity atom numbers $N$ (see legend of the Figure).
We observe that $g^{(2)}(0)$ rapidly changes from $g^{(2)}(0)=2$ (chaotic light) to $g^{(2)}(0)=1$ (coherent light) at the threshold between the non-superradiant and the superradiant phase. This behavior can be understood since well inside the non-superradiant phase $\hat{J}^{+}$ and $\hat{J}^{-}$ are purely noisy and therefore we have $\langle \hat{J}^{+}\hat{J}^{+}\hat{J}^{-}\hat{J}^{-}\rangle\approx2\langle \hat{J}^{+}\hat{J}^{-}\rangle^2$ implying $g^{(2)}(0)\approx2$. On the other hand, well inside the superradiant regime the collective dipole for $N\to\infty$ is determined by its mean-field value $\langle \hat{J}^{+}\hat{J}^{+}\hat{J}^{-}\hat{J}^{-}\rangle\approx\langle \hat{J}^{+}\hat{J}^{-}\rangle^2\approx J_{\mathrm{st}}^2/4$. This gives $g^{(2)}(0)\approx1$. This claim is supported by the simulations with different intracavity atom numbers that show that for larger $N$ the values of $g^{(2)}(0)$ seem to converge to 1. The fact that $g^{(2)}(0)$ is increasing for larger values of $\Phi\tau^2\Gamma_c$ is consistent with the fact that the mean-field value of the collective dipole $J_{\mathrm{st}}$ (see Fig.~3 in the main text) is decreasing with increasing $\Phi\tau^2\Gamma_c$. Consequently, fluctuations introduced by noise are more pronounced with respect to the mean-field value of the collective dipole and give rise to the deviation from $g^{(2)}(0)=1$.

We want to highlight that $g^{(2)}(0) \approx 1$ for $\Phi\tau^2\Gamma_c = 20$ and $\delta_D\tau \lesssim 0.2\pi$  implies that in vicinity of the optimal operational regime the superradiant beam laser is second-order coherent. This is important since this is the region where the laser operates with near maximum output power and near minimum linewidth.

\section{Environmental Noise\label{sec: cavityNoise}}
\subsection{Effective acceleration sensitivity, cavity length, and coefficient of thermal expansion}

In this section, we show how cavity pulling results in an effective acceleration sensitivity $\wp K$. We also show that, due to cavity pulling, thermal fluctuations behave as if the cavity has an effective length $L/\wp$.

Classically, the electric field of a laser with a finite linewidth is

\begin{equation}
    E(t) = E_0 e^{i\left[\omega_0 t + \theta(t)\right]},
\end{equation}

\noindent where $E_0$ is the (constant) electric field amplitude, $\omega_0$ is the center frequency of the laser, and $\theta(t)$ is the laser phase, which is a stochastic function that is responsible for linewidth broadening.

For a cavity-stabilized light source, the laser is tightly locked to the peak of a cavity resonance; therefore, $\theta(t) = \phi(t)$, where $\phi(t)$ is the phase noise due to disturbances in the center of the cavity fringes (caused by environmental noise). The laser linewidth is given by the full width at half maximum of the laser's power spectral density $S_E(\omega)~$\cite{didomenico2010}, which for a cavity-stabilized laser is

\begin{equation}
    S_E(\omega)=2 E_0^2 \int_{-\infty}^{\infty} d\tau \, e^{-i (\omega - \omega_0)\tau} e^{-\frac12 \langle \left[ \phi(t+\tau) - \phi(t) \right]^2 \rangle}.
\end{equation}

\noindent Here $\langle \ldots \rangle$ denotes a time average.

In the superradiant beam laser, the phase noise of the output electric field is $\theta(t) = \wp \phi(t)$, which is suppressed by a factor of $\wp<1$ (compared to a cavity-stabilized laser) due to the atom-cavity interaction. Consequently, the power spectral density of the beam laser output is

\begin{equation}
    \label{eqn:psd_beam_laser}
    S_E(\omega)=2 E_0^2 \int_{-\infty}^{\infty} d\tau \, e^{-i (\omega - \omega_0)\tau} e^{-\frac12\wp^2\langle \left[ \phi(t+\tau) - \phi(t) \right]^2 \rangle}.
\end{equation}
The time average of the phase can be evaluated with the expression \cite{riehle2003}

\begin{equation}
    \frac12 \langle \left[ \phi(t+\tau) - \phi(t) \right]^2 \rangle = 2 \int_0^{\infty} df \, S(f) \frac{\sin^2(\pi f \tau)}{f^2},
\end{equation}

\noindent where $S(f)$ is the noise spectral density of the laser. 

For the case of vibration noise in 1 dimension, $S(f) = K^2 \, G_a(f)$, where $K$ is the acceleration sensitivity and $G_a(f)$ is the spectrum of accelerations due to mechanical vibration \cite{filler1988}. Due to the presence of $\wp^2$ in Eq.~\eqref{eqn:psd_beam_laser}, the output behaves as if it has an effective vibration noise spectral density $K_{\mathrm{eff}}^2 G_a(f)$, where

\begin{equation}
    K_{\mathrm{eff}} = \wp K.
\end{equation}

Meanwhile for thermal Brownian noise, $S(f) = G_L(f)/L^2$, where $L$ is the cavity length and $G_L(f)$ is the spectral density of length fluctuations \cite{martin2012}. As in the case of vibration noise, the beam laser output has an effective thermal noise spectral density $G_L(f)/L_{\mathrm{eff}}^2$, where

\begin{equation}
    L_{\mathrm{eff}} = \frac{L}{\wp}.
\end{equation}

Lastly, we consider slow thermal drift of the cavity length. For a cavity-stabilized laser, the output frequency $\omega$ is equal to the cavity resonance frequency $q c/2 L$, where $q$ is the mode order. When the temperature changes, we find that $d\omega/dT = -\alpha \omega_c$, where $\alpha = \frac{1}{L} \frac{dL}{dT}$ is the coefficient of thermal expansion (CTE) of the cavity. 

For the beam laser, we use the definition of cavity pulling, $\wp = (\omega - \omega_a)/(\omega_c - \omega_a)$, and find that $d\omega/dT = -\omega_c \wp \alpha$. Therefore, the beam laser has an effective CTE of 

\begin{equation}
    \alpha_{\mathrm{eff}} = \wp \, \alpha.
\end{equation}

\subsection{Sensitivity to environmental noise}

Ultrastable cavities used to generate narrow lasers are carefully engineered to be insensitive to vibrations. They are mounted with vibration cancelling schemes, housed inside acoustic shields, and set up on vibration damping platforms. The record best cavity-stabilized laser has a vibration sensitivity on the order of $K \sim 10^{-13}/(\mathrm{m/s^2})$ \cite{robinson2019}.

Meanwhile if one were to design a simple cavity mounted on a V block and ignore vibration isolation completely, the vibration sensitivity of the cavity resonance would be on the order of $K \sim 10^{-9}/(\mathrm{m/s^2})$ \cite{chen2006}. However, in the case of the superradiant beam laser, because of cavity pulling, the effect of this vibration on the laser output is given by $\wp K \sim 10^{-13}/(\mathrm{m/s^2})$, where $\wp = 0.004$ (as in the main text). Therefore, with regard to environmental noise, a minimalist cavity in a beam laser could be competitive with the record best ultrastable cavity achieved to date.

A common approach to reducing thermal noise is to operate with longer cavities, which have been made as long as half a meter \cite{hafner2015}. It may be difficult to make cavities considerably longer because larger cavities have greater acceleration sensitivity. However, a compact 3 cm long cavity used in a superradiant beam laser has the thermal noise of a cavity with the effective length $L/\wp = 7.5\,\mathrm{m}$.

In addition to vibration and thermal noise, many cavities deal with drift on long timescales. A major contributor to drift is thermal expansion of the cavity when its temperature changes. To determine the thermal expansion, a good model of the CTE is

\begin{equation}
    \alpha = A (T-T_0),
\end{equation}

\noindent where $A \sim 1 \times 10^{-9} \, \mathrm{K}^{-2}$, and $T_0$ is the zero crossing temperature \cite{legero2010}. In ultralow expansion glass (ULE) cavities, $T_0$ is typically engineered to be near room temperature, and the cavity is then temperature controlled to $T_0$. However, even a 10 $\mu$K fluctuation away from $T_0$ causes a change in the cavity resonance of $-A(\omega_c/2\pi)T \times (10\,\mathrm{\mu K}) = 1.5\,\mathrm{kHz}$ (where $T = 300 \, \mathrm{K}$), which is a challenge because it is orders of magnitude larger than the laser linewidth.

Cavities for simpler applications often use Invar, which is easier to work with than ULE. For small fluctuations about room temperature, Invar has a CTE of $\alpha \sim 1 \times 10^{-6} \, \mathrm{K}^{-1}$. This is treated as constant since it does not have a zero crossing at room temperature. Standard temperature control can achieve 1 mK level stability without the addition of heat shielding; therefore, an Invar cavity that is modestly temperature controlled can achieve the same thermal drift rates as highly temperature stabilized ULE cavities.

\section{Sample Design Constraints\label{sec: designConstraints}}
In this section, we give two examples of sample experimental parameters that are consistent with the production of ultranarrow linewidth laser light within our presented theoretical framework. We have chosen as our examples the $2\pi \times 400\, \mathrm{Hz}$ transition line of $^{40} \mathrm{Ca}$ and the $2\pi \times 7.5\, \mathrm{kHz}$ transition line of $^{88} \mathrm{Sr}$, respectively. In order to emphasize the potential simplicity, we consider here the most direct implementation, i.e., a hot atomic beam, a single mode cavity, and a simplified model of the atom delivery system where transverse collimation or transverse laser cooling is implemented but no longitudinal cooling is assumed. On the other hand, we recognize that in a real device, transverse cooling, velocity selective techniques, or alternative beam delivery approaches, might be employed in order to more easily satisfy the Doppler constraint and enhance the effective atomic beam flux through cavity mode. In the case of a more sophisticated design choice, the parameters may be considerably more favorable than the numbers we give here for a simple configuration. However, the constraints $\Phi\tau^2\gc > 8$ and $\delta_D \tau = k \Delta v_z \tau < \pi$ must always be satisfied in order to realize CW superradiant emission with linewidth in the ultranarrow regime, which is a principal point of our proposal.

\subsection{First Example: $2 \pi \times 400 \, \mathrm{Hz}$ Line of ${}^{40} \mathrm{Ca}$\label{sec: Ca}}
\begin{table}[h]
\begin{center}
\begin{tabular}{|c|c|c|}
\hline
Transition Rate  & $\gamma$ & $2 \pi \times (400 \, \mathrm{Hz}$)\\
\hline
Effective Beam Rate & $\Phi$ & $6.1 \times 10 ^ {14}$/s \\
\hline
Transit Time & $\tau = \frac{2w}{v_L}$ &  0.81 $\mu$s \\
\hline
Intracavity Atom Number & $N=\Phi \tau$ & $4.9 \times 10^8$\\
\hline
Transverse Velocity Threshold & $\Delta v = \frac{\lambda}{2 \tau}$  & 41 cm/s\\
\hline
Minimum Linewidth & $\gc = \gamma \mathcal{C}$ & $2 \pi \times (8 \, \mathrm{mHz}$) \\
\hline
Peak Power & $P_{max} \approx 0.7 \Phi \hbar \omega$ &  $0.1 \, \mathrm{mW}$\\
\hline
Cavity Pulling & $\wp$ &  $0.004$\\
\hline
\end{tabular}
\end{center}
\caption{Sample model parameters for the superradiant beam laser utilizing the $2\pi \times 400\, \mathrm{Hz}$ transition line of $^{40} \mathrm{Ca}$.}
\label{table: eg1}
\end{table}

For the Ca case (see Table~\ref{table: eg1}), we consider a hot oven operating at temperature $842$~\textdegree{}C, which gives an out-of-oven beam rate $\Phi_0 \sim 10^{19}\,\mathrm{/s}$. As the atomic beam propagates from the oven to the cavity, we assume that no longitudinal cooling is used, so the mean longitudinal velocity is well approximated by the mean Maxwellian velocity as $v_L = 765.9\,\mathrm{m/s}$. In the transverse direction, laser cooling or velocity selection is needed to restrict the transverse Doppler width to below $\delta_D = 2\pi \times 0.6\, \mathrm{MHz}\,(\Delta v_z = 0.41\,\mathrm{m/s})$, in order to operate below the critical Doppler threshold discussed in the main text. Prior to the atomic beam entering the cavity, the atoms must be optically pumped into the excited electronic state. We estimate that for this situation, the effective beam rate entering the cavity should be of order $\Phi = 6.1 \times 10 ^ {14}$/s. We choose cavity parameters by considering a lossy cavity of length $L=3.3\, \mathrm{cm}$, beam waist $w=0.31\, \mathrm{mm}$, and finesse $F=22.8$, which corresponds to a cavity decay $\kappa = 2 \pi \times (197\, \mathrm{MHz})$ and cavity cooperativity $\mathcal{C} = 2 \times 10^{-5}$. Given these parameters, our calculation predicts an output field of power $0.1\, \mathrm{mW}$ and linewidth of the order $10\, \mathrm{mHz}$. Meanwhile, $\kappa\tau \approx 1000$ gives a cavity pulling $\wp$ as small as 0.004 as shown in Fig.~4 of the main text. 

\subsection{Second Example: $2\pi \times 7.5\, \mathrm{kHz}$ Line of $^{88}\mathrm{Sr}$ \label{sec: Sr}}
\begin{table}[h]
\begin{center}
\begin{tabular}{|c|c|c|}
\hline
Transition Rate  & $\gamma$ & $2\pi \times (7.5\, \mathrm{kHz})$\\
\hline
Effective Beam Rate & $\Phi$ & $1.2 \times 10 ^ {13}$/s\\
\hline
Transit Time & $\tau$ &  1.3 $\mu$s\\
\hline
Intracavity Atom Number & $N$ & $1.6 \times 10^7$\\
\hline
Transverse Velocity Threshold & $\Delta v$ & 26 cm/s\\
\hline
Minimum Linewidth & $\gc$ & $2 \pi \times (150 \, \mathrm{mHz}$)\\
\hline
Peak Power & $P_{max}$& $2.5 \, \mathrm{\mu W}$\\
\hline
Cavity Pulling & $\wp$ & $0.004$\\
\hline
\end{tabular}
\end{center}
\caption{Sample model parameters for the superradiant beam laser utilizing the $2\pi \times 7.5\, \mathrm{kHz}$ transition line of $^{88} \mathrm{Sr}$.}
\label{table: eg2}
\end{table}

For the Sr case (see Table~\ref{table: eg2}), we consider a similar experimental design to the case for Ca. The oven operating at $650$~\textdegree{}C gives a $\Phi_0 \sim 10^{18}\,\mathrm{/s}$ and $v_L = 469.8\,\mathrm{m/s}$. The required Doppler threshold is $\delta_D = 2\pi \times 0.4\, \mathrm{MHz}\,(\Delta v_z = 0.26\,\mathrm{m/s})$, and the required effective beam rate is $\Phi = 1.2 \times 10 ^ {13}$/s. Considering a cavity of length $L=6.0\, \mathrm{cm}$, beam waist $w=0.31\, \mathrm{mm}$, and finesse $F=20.8$, which corresponds to $\kappa = 2 \pi \times (121\, \mathrm{MHz})$ and cavity cooperativity $\mathcal{C} = 2 \times 10^{-5}$, we calculate an output field of power $2.5\, \mathrm{\mu W}$ and linewidth of the order $150\, \mathrm{mHz}$. Since $\kappa\tau \approx 1000$, we predict a cavity pulling coefficient of $\wp \approx 0.004$.